\RequirePackage{etoolbox}
\PassOptionsToClass{letterpaper}{article} % arXiv look

\PassOptionsToPackage{hypertexnames=false}{hyperref}
\RequirePackage{silence}
\newif\iffeyn
\feyntrue
\documentclass[submission, Phys]{SciPost}

\usepackage{xspace}

\hypersetup{
    colorlinks,
    linkcolor={red!50!black},
    citecolor={blue!50!black},
    urlcolor={blue!80!black}
}

\usepackage[bitstream-charter]{mathdesign}
\DeclareSymbolFont{usualmathcal}{OMS}{cmsy}{m}{n}
\DeclareSymbolFontAlphabet{\mathcal}{usualmathcal}

\usepackage{XCharter}

\allowdisplaybreaks

\usepackage{bm,bbm}
\usepackage{enumitem}
\usepackage[normalem]{ulem}
\usepackage{colortbl}
\usepackage{cancel}

\usepackage{empheq}

\usepackage{amsmath}

\iffeyn
\usepackage[compat=1.1.0]{tikz-feynman} 
\fi

  \usepackage{float}
  \usepackage{afterpage}
  \usepackage{array}
  \usepackage{booktabs}
  \usepackage{multirow}
\newcommand{\fref}[1]{\hyperref[#1]{Figure~\ref*{#1}}}
\newcommand{\Fref}[1]{\hyperref[#1]{Figure~\ref*{#1}}}
\newcommand{\sref}[1]{\hyperref[#1]{Section~\ref*{#1}}}
\newcommand{\Sref}[1]{\hyperref[#1]{Section~\ref*{#1}}}
\newcommand{\tref}[1]{\hyperref[#1]{Table~\ref*{#1}}}
\newcommand{\Tref}[1]{\hyperref[#1]{Table~\ref*{#1}}}
\newcommand{\aref}[1]{\hyperref[#1]{Appendix~\ref*{#1}}}
\newcommand{\Aref}[1]{\hyperref[#1]{Appendix~\ref*{#1}}}
\newcommand{\thref}[1]{\hyperref[#1]{Theorem~\ref*{#1}}}
\newcommand{\Thref}[1]{\hyperref[#1]{Theorem~\ref*{#1}}}
\newcommand{\alref}[1]{\hyperref[#1]{Algorithm~\ref*{#1}}}
\newcommand{\Alref}[1]{\hyperref[#1]{Algorithm~\ref*{#1}}}

\newcommand{\pye}{\textsc{Pythia8}\xspace}
\newcommand{\pwgb}{\textsc{POWHEG-BOX}\xspace}
\newcommand{\amc}{\textsc{MadGraph5\_aMC@NLO}\xspace}
\newcommand{\amcs}{\textsc{MG5\_aMC}\xspace}
\newcommand{\hws}{\textsc{Herwig7}\xspace}

\newcommand{\bt}{\bar{t}}
\newcommand{\bq}{\bar{q}}
\newcommand{\pt}{p_{\textsc{T}}}
\newcommand{\kt}{k_{\sss T}}

\renewcommand{\Theta}{{\bm{\theta}}}

\renewcommand{\l}{\bm{l}}

\newcommand{\beq}{\begin{equation}}
\newcommand{\eeq}{\end{equation}}
\newcommand{\sss}{\sc}

\graphicspath{{figures/}}

\newcommand{\reportnumber}{FERMILAB-PUB-23-413-CSAID}
\usepackage[angle=0,scale=1,color=black,firstpage=true,opacity=1]{background}

\SetBgContents{\footnotesize\sf{\reportnumber}}
\SetBgPosition{current page.north east}
\SetBgHshift{-2in}
\SetBgVshift{-0.5in}
\begin{document}

\begin{center}{\Large \textbf{
Matrix element corrections in the \textsc{Pythia8} parton shower in the context
of matched simulations at next-to-leading order 
}}\end{center}

\begin{center}
Stefano Frixione\textsuperscript{1},
Simone Amoroso \textsuperscript{2},
Stephen Mrenna\textsuperscript{3}
\end{center}

\begin{center}
{\bf \textsuperscript{1}} {INFN, Sezione di Genova, Via Dodecaneso 33, 
I-16146, Genoa, Italy}
\\
{\bf \textsuperscript{2}} {Deutsches Elektronen-Synchrotron Notkestr. 85, 
22607 Hamburg, Germany}
\\
{\bf \textsuperscript{3}}~Scientific~Computing~Division,~Fermi~National~Accelerator~Laboratory,~Batavia,~IL~60510,~USA
\\
\end{center}

\begin{center}
\today
\end{center}

% For convenience during refereeing (optional),
% you can turn on line numbers by uncommenting the next line:
%\linenumbers
% You should run LaTeX twice in order for the line numbers to appear.

\section*{Abstract}
{We discuss the role of matrix element corrections (MEC) to parton 
showers in the context of MC@NLO-type matchings for processes that feature
unstable resonances, where MEC are liable to result in double-counting 
issues, and are thus generally not employed. By working with \textsc{Pythia8}, 
we show that disabling all MEC is actually unnecessary in computations 
based on the narrow-width approximation, and we propose alternative MEC
settings which, while still avoiding double counting, allow one to
include hard-recoil effects in the simulations of resonance decays.
We illustrate our findings by considering $t\bar{t}$ production at
the LHC, and by comparing \textsc{MadGraph5\_aMC@NLO} predictions
with those of \textsc{POWHEG-BOX} and standalone \textsc{Pythia8}.}

\vspace{10pt}
\noindent\rule{\textwidth}{1pt}
\tableofcontents\thispagestyle{fancy}
\vspace{10pt}

\section{Introduction\label{sec:intro}}
The exclusive simulation of processes that feature heavy unstable 
particles, such as the top, $W$, $Z$, and Higgs in the Standard Model,
is complicated for a variety of reasons. Conceptually, the most
straightforward approach is to choose a (set of)
decay channel(s) for the unstable particle(s) and compute the
process where the initial and final states only include light
partons and the products of such decay channels. For example,
in the case of $t\bar{t}$ production, one may focus on a dilepton
channel, and thus simulate (in hadronic collisions)\footnote{We shall
use the case of $t\bar{t}$ production in all of our examples; it should
be clear that the arguments are general and apply to any process.}
\beq
pp\;\longrightarrow\;b\bar{\ell}_i\nu_i\bar{b}\ell_j\bar\nu_j\,,
\label{ttfull}
\eeq
rather than
\beq
pp\;\longrightarrow\;t\bar{t}\,.
\label{ttred}
\eeq
From the viewpoint of computing resources, the process of eq.~(\ref{ttfull})
is much more expensive than that of eq.~(\ref{ttred}). This does not
change when one takes into account the fact that eq.~(\ref{ttred}) must
be supplemented by the simulation of the leptonic decays of the tops,
namely:
\beq
t\;\longrightarrow\;b\bar{\ell}_i\nu_i\,,\;\;\;\;\;\;\;\;
\bar{t}\;\longrightarrow\;\bar{b}\ell_j\bar\nu_j\,,
\label{tdecs}
\eeq
since the complexity of such a simulation is generally smaller than
that of either eq.~(\ref{ttfull}) or~(\ref{ttred})\footnote{A more
extended discussion of the arguments that follow can be found
e.g.~in sect.~2.5 of ref.~\cite{Alwall:2014hca}.}. Having said that,
one remarks that the combination of eqs.~(\ref{ttred}) and~(\ref{tdecs})
is equivalent to eq.~(\ref{ttfull}) only in the limit of a vanishing
top-quark width (the so-called narrow-width approximation); away from
that limit, the agreement between the results of the two approaches
might be degraded by the presence of non-resonant (i.e., non-$t\bar{t}$
mediated) contributions. Even when only resonant contributions are
present, it is generally the case that the incoherent simulation of
the production process (eq.~(\ref{ttred})) and of the decay processes
(eq.~(\ref{tdecs})) constitutes a much poorer physics description
than that emerging from eq.~(\ref{ttfull}), and this because of
two mechanisms. Firstly, the kinematics of a decay product of
a given unstable particle may be affected by that of partons/leptons
in the event which are themselves not decay products of that unstable
particle. These effects are called (production) spin correlations,
and cannot be accounted for by the separate simulations of the
individual decays, while they are correctly included by the matrix
elements that underpin eq.~(\ref{ttfull}). Secondly, the kinematics
of the decay products is affected by the emissions of the extra
partons that appear in the hard process when perturbative corrections
are considered, as in the case of NLO+PS simulations. By construction,
a parton shower mimics the effects of such extra emissions, however
with an accuracy which decreases with the hardness of the recoils
they induce. Therefore, while perturbative corrections to eq.~(\ref{ttfull})
will automatically include hard-recoil effects, such effects are
poorly described when the simulation of the decays of eq.~(\ref{tdecs}) is
followed by an ordinary parton shower.

As far as the first mechanism is concerned, spin correlations are
nowadays routinely taken into account in the context of production+decay
simulations (eqs.~(\ref{ttred}) and~(\ref{tdecs})) by means of the
procedures of ref.~\cite{Frixione:2007zp} (especially in NLO+PS
approaches) or ref.~\cite{Richardson:2001df}; we shall not discuss
them any further in this work. Conversely, hard recoils, absent
computations of the complete processes such as that of eq.~(\ref{ttfull}),
are approximated by Monte Carlo event generators (MC henceforth) by means 
of the so-called
Matrix Element Corrections (MEC henceforth). The manner in which MEC are
simulated depends on the specific MC one employs, but they all rely
on (exact or approximated) tree-level matrix elements that describe
the emission(s) of interest.
%%%%%%%%%%%%%%%%%%%%%%%%%%%%%%%%%%%%%%%%%%%%%%%%%%%%%%%%%%%%%%%%%%%%%%%%
\iffeyn
\begin{figure}[!htb]
  \centering
    \resizebox{0.4\textwidth}{!}{%
    \begin{tikzpicture}
      \begin{feynman}
        \vertex (a1) ;
        \vertex[right=1.5cm of a1] (a2);
        \vertex[right=1.0cm of a2] (a3);
        \vertex[right=1.5cm of a3] (a4);
        \vertex at ($(a1)!0.75!-20:(a4)$) (d);
        \vertex[above=2em of a4] (c1) ;
        \vertex[above=2em of c1] (c3);
        \vertex at ($(c1)!0.5!(c3) - (1cm, 0)$) (c2);
        \diagram* {
          (a1) -- [fermion,very thick] (a2) -- [fermion,very thick] (a3) -- [fermion] (a4),
          (c3) -- [fermion, out=180, in=45] (c2) -- [fermion, out=-45, in=180] (c1),
          (a3) -- [boson,bend left] (c2),
          (a2) -- [gluon] (d)
        };
        \node[draw,fill=blue!20!white,circle,minimum size=0.05\textwidth] at (a1) {\(\cal H\)};
      \end{feynman}
    \end{tikzpicture}
  }
  \resizebox{0.4\textwidth}{!}{%
    \begin{tikzpicture}
      \begin{feynman}
        \vertex (a1) ;
        \vertex[right=1.5cm of a1] (a2);
        \vertex[right=1cm of a2] (a3);
        \vertex[right=1.5cm of a3] (a4);
        \vertex at ($(a1)!0.95!-15:(a4)$) (d);
        \vertex[above=2em of a4] (c1) ;
        \vertex[above=2em of c1] (c3) ;
        \vertex at ($(c1)!0.5!(c3) - (1cm, 0)$) (c2);
        \diagram* {
          (a1) -- [fermion,very thick] (a2) -- [fermion] (a3) -- [fermion] (a4),
          (c3) -- [fermion, out=180, in=45] (c2) -- [fermion, out=-45, in=180] (c1),
          (a2) -- [boson, bend left] (c2),
          (a3) -- [gluon] (d)
        };
        \node[draw,fill=blue!20!white,circle,minimum size=0.05\textwidth] at (a1) {\(\cal H\)};
      \end{feynman}
    \end{tikzpicture}
  }
  \caption{\label{fig:MECex}
     Sample graphs relevant to MEC in top production (left panel) and top 
     decay (right panel). The top quark is depicted by means of a thicker 
     fermion line.
     The hard production process $\cal H$ is indicated by the blue circle.
  }
\end{figure}
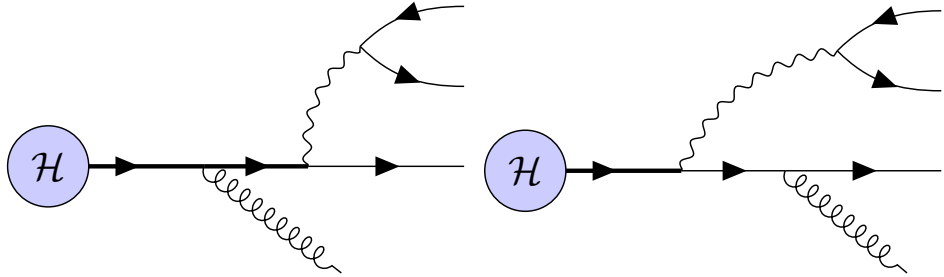
%%%%%%%%%%%%%%%%%%%%%%%%%%%%%%%%%%%%%%%%%%%%%%%%%%%%%%%%%%%%%%%%%%%%%%%%
\else
\begin{figure}[!htb]
\centering
 \includegraphics[width=.4\textwidth]{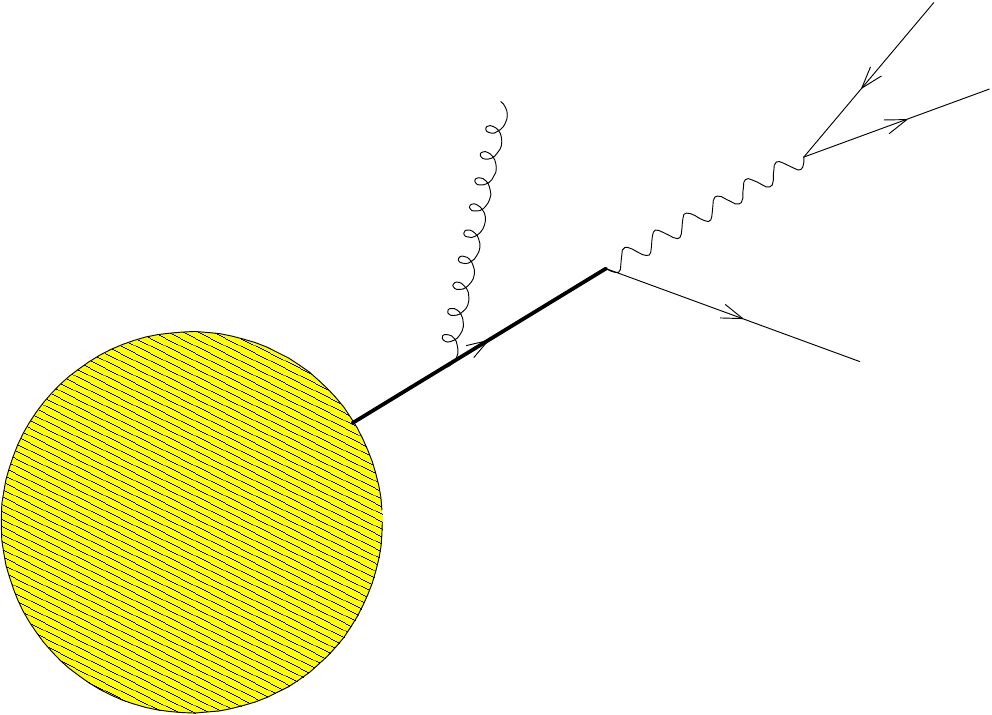}
$\phantom{aaa}$
 \includegraphics[width=.4\textwidth]{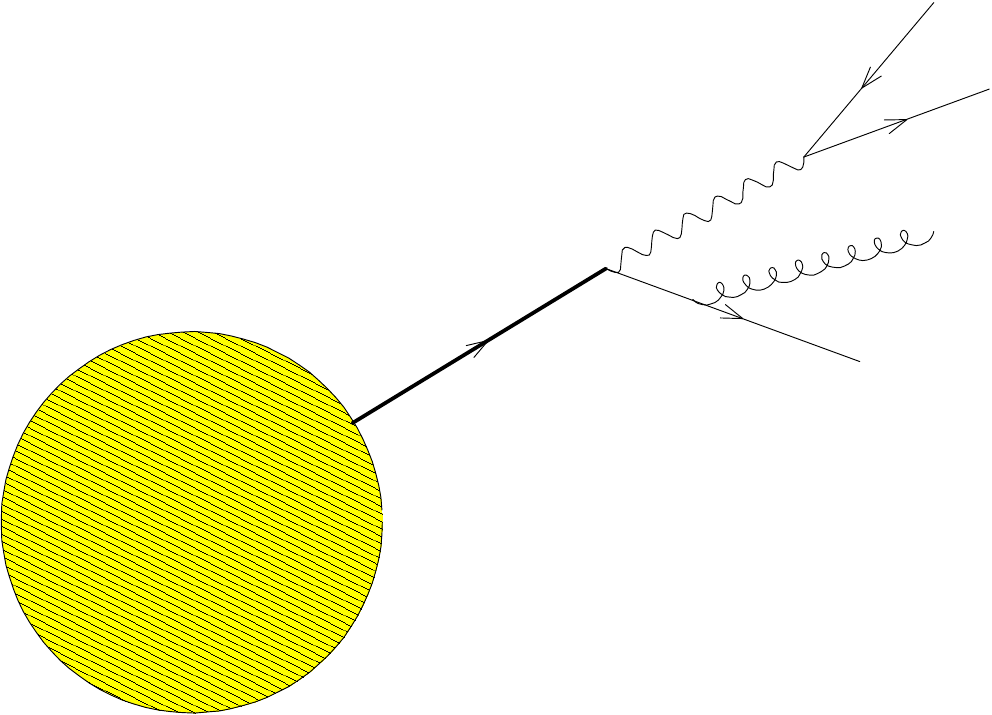}
 \caption{\label{fig:MECex}
Sample graphs relevant to MEC in top production (left panel) and top 
decay (right panel). The top quark is depicted by means of a thicker 
fermion line.}
\end{figure}

\fi

Before the widespread adoption of matching and merging techniques,
MCs (and specifically Pythia) used to add MEC to both the production 
(eq.~(\ref{ttred})) and the decay (eq.~(\ref{tdecs})) processes.
These mechanisms will henceforth be referred to as production and
decay MEC, respectively; sample Feynman graphs involved in their
computations, in the specific case of top-quark production and
decay, are depicted in fig.~\ref{fig:MECex}. It must be stressed
that MEC graphs for production and decay are {\em not} allowed to
interfere; in other words, they are employed incoherently.

In the context of an NLO+PS simulation\footnote{Or of a
multijet-merged one; for what we are concerned with here, the two
are identical, and we shall refer to the former while generally
understanding also the latter.}, MEC can potentially spoil the
accuracy of the computation by double counting (essentially, by generating 
again an emission already accounted for at the hard matrix element level -- 
this is the case for the contribution shown on the left-hand 
panel of fig.~\ref{fig:MECex}). In fact, this is not an issue 
for decay MEC, even in the presence of spin-correlation
corrections, since the latter do not feature (at present) any extra
emissions. There is therefore a good motivation to always include decay 
MEC in one's simulation, since they may induce, and accurately account 
for, large shifts in observables sensitive to emissions off decay
products\footnote{For an example relevant to top-mass determinations,
see e.g.~\cite{FerrarioRavasio:2018whr}.}. In contrast, the case of
production MEC must be considered more carefully. In 
MC@NLO-type~\cite{Frixione:2002ik}
simulations, the MC counterterms that guarantee the absence of
double counting are determined by considering the showers {\em without}
any additional MEC; this implies that {\em production} MEC are certain to
spoil the accuracy of an MC@NLO simulation. This is not a simple formal
problem; since emissions in MC@NLO are not ordered in hardness, it is
not unlikely that the effects of such double counting can be visible
in phenomenologically-relevant quantities. On the other hand, in a
POWHEG-type simulation the hardest emission is always generated before
the MC shower is added; potential NLO effects driven by production
MEC are thus restricted to the region where the POWHEG-generated emission
did not occur, i.e.~a small-$\pt$ region where the cross section
is Sudakov suppressed.

The bottom line is that both MC@NLO- and POWHEG-based simulations
are compatible with, and phenomenologically benefit from, decay MEC.
Conversely, while production-MEC effects are essentially negligible
in POWHEG and can be either included or discarded, they are responsible
for double counting in MC@NLO, and must therefore not be employed there.
Thus, the procedure to adopt in MC@NLO and POWHEG is in principle
clear. In practice, unfortunately, the MC@NLO-type MC interfaces 
to date have disabled \textit{all} MEC instead 
of just those necessary to avoid double counting.
We have already mentioned that the phenomenological implications of
such choices should be restricted to a certain narrow class of observables.
However, observed differences between MC@NLO- and POWHEG-based $t\bar{t}$ 
predictions at the LHC are commonly attributed to the different underlying 
matching mechanisms, whereas it might be the case, and certainly for
the observables mentioned above, that they stem from the absence
as opposed to the presence, respectively, of decay-MEC effects.

The aim of this note is to document how the separation of production and decay
MEC can be achieved\footnote{To the best of our understanding, this separation
is also possible in \hws~\cite{Bellm:2015jjp}.} in 
\pye~\cite{Bierlich:2022pfr}, thus paving the way for MC@NLO-based simulations
which are phenomenologically more complete than those carried out thus far. We
shall present examples for $t\bar{t}$ production, where \pye standalone runs
are compared with POWHEG-based simulations performed with
\pwgb~\cite{Alioli:2010xd}, and MC@NLO-based simulations performed with
\amc~\cite{Alwall:2014hca} (shortened as \amcs henceforth).

\section{Matrix Element Corrections in the \pye dipole shower
\label{sec:py8MEC}}
An introduction to the basics of MEC has been given in sect.~\ref{sec:intro};
here, we focus on their implementations within \pye, and explicitly
consider the various settings that must be chosen in order to preserve the 
accuracy and precision of the underlying perturbative predictions.
The \pye standalone parton shower algorithm is based on a leading-logarithm 
DGLAP evolution, which is by default corrected by means of MEC to reproduce 
the appropriate tree-level matrix element behaviour, including in the 
cases where \pye is interfaced with external calculations. Technically,
this is done by populating all of the phase space by candidate emissions
whose number is overestimated, and which are then possibly vetoed with a
rejection rate given by the ratio of the matrix elements over their
shower approximations.  While the MEC provide continuity between the exact 
matrix-element based calculation and the parton shower in the hard emission 
limit, it also provides the soft emission limit necessary
for particles with mass on the order of the soft emission scale.
Various options are available that allow for all, some, or none of the MEC to
be included in a given simulation. It is therefore important to understand
which of these options need to be employed in the context of NLO+PS
simulations, so as to avoid the double counting issue discussed in
sect.~\ref{sec:intro}.

We start by pointing out that the separation between MEC in production
and decay, exemplified by the two graphs in fig.~\ref{fig:MECex},
is conceptually well defined in the zero-width limit (the width being
that of the particle that decays), since in such a limit the interference
between the two types of tree-level graphs vanishes; this is important,
because MEC are effected at the level of amplitude squared. However, this
is not the way in which MCs, and specifically \pye, operate, since the
primary distinction there is that between emissions off initial-state
and off final-state legs (conventionally referred to as ISR and FSR,
respectively). Thus, production MEC are relevant to both ISR and FSR,
whereas decay MEC are solely relevant to FSR. This implies that production
MEC constitute a difficult problem, since in general at the level of amplitude
squared one cannot unambiguously separate the effects due the graphs
where the extra parton is emitted by an initial-state leg from those
due to emissions by a final-state leg\footnote{Another complication
is that of a final-state configuration that can be produced by means
of several partonic channels. For example, the matrix element of
$qg\to t\bt q$ would be relevant to {\em both} a $q\to qg$ branching
attached to $gg\to t\bt$ and a $g\to q\bq$ branching attached to 
$q\bq\to t\bt$.}. 

The \pye solution to the problem posed by production MEC is rather
draconian. Namely, MEC are applied to ISR only for colour-singlet
resonant production~\cite{Miu:1998ju} (i.e.~to $q\bq\to V$, with $V$ 
a heavy EW boson, and to $gg/\gamma\gamma\to H$), since in such a case 
the ambiguity mentioned before is absent. As far as FSR is concerned,
one applies MEC by considering, for each possible emitting colour dipole,
a matrix element deemed equivalent to it, according to table~1
of ref.~\cite{Norrbin:2000uu} (for example, for $t\bt$ production,
the relevant matrix element is $\gamma^*\to t\bt g$).

According to what was said previously, the latter FSR mechanism applies
to decay MEC as well. However, before resorting to that, priority is
given to matrix elements which {\em exactly} match the decay one
is considering\footnote{Note that the same strategy would apply
to production MEC if exactly-matching matrix elements could be found
with an unambiguous separation between ISR and FSR, of which there is
none that cannot also be seen as a production+decay mechanism
(e.g.~$q\bq\to Z^\star\to q\bq$; see later).}; if those exist, they are used 
for the MEC. In the Standard Model, these include the $t\to Wbg$ and
the $V\to q\bq$ matrix elements; in particular, this implies that
a top that decays hadronically is never seen as such, but rather as
a decay chain of a top decaying to a $Wb$ pair, followed by a
hadronic $W$ decay.

Regardless of whether equivalent or exactly-matching matrix elements
are employed for MEC, their effects can be iterated by \pye for subsequent
emissions as well (as opposed to limiting them to the first emission).
This is done by applying MEC on the system obtained by discarding
all radiated partons except that whose kinematical configuration one
seeks to correct presently; this may require a momentum reshuffling 
(for example: if a $t\to Wbgg$ configuration is obtained by means of
two subsequent emissions off the $b$, such emissions can be both corrected,
one after the other, by using the $t\to Wbg$ matrix element; $g\to gg$ 
branchings are not corrected). For a decay chain (such as $t\to W(\to q\bq)b$),
the iteration of MEC is applied individually to each resonance that
decays (here, the top and the $W$).

The various strategies described above are controlled in \pye by means
of the following parameters, separately for ISR (\texttt{X=SpaceShower})
and FSR (\texttt{X=TimeShower}), which are to be set equal to either
\texttt{on} (which is the default for all of them) or \texttt{off}:
\begin{description}
\item[\texttt{X::MEcorrections}]: If \texttt{=on}, use MEC whenever 
available, regardless of whether they entail the use of equivalent 
or exactly-matching matrix elements.
\item[\texttt{X::MEextended}]: If \texttt{=on}, use equivalent matrix 
elements for MEC;\\
ignored if \texttt{X::MEcorrections=off}.
 \item[\texttt{X::MEafterFirst}]: If \texttt{=on}, apply MEC also to all 
emissions after the first;\\ ignored if \texttt{X::MEcorrections=off}.
\end{description}
Among other things, the above implies that if \texttt{X::MEcorrections=on},
then the MEC stemming from exactly-matching matrix elements are always
applied, irrespective of the value of \texttt{X::MEextended}.
Thus, presently one can generally switch production MEC off while
keeping decay MEC on, but not the other way around. This is just as
well, since as it was discussed in sect.~\ref{sec:intro} MEC in production
have to be employed only when neither NLO matching nor multi-jet merged
simulations are available for the process of interest (which nowadays is
a virtual impossibility), whereas MEC in decay are important for
better phenomenological predictions in the context of relatively
cheap (CPU-wise) generations. Having said that, this discussion clarifies
that a more flexible approach to MEC in \pye with respect to that given by the
three (per shower type) parameters above would be desirable, namely one that
allows the MEC to be applied or disabled until a certain condition is
met\footnote{See the discussion on this point in appendix~\ref{app:c}.}.

In summary, the necessity of avoiding double counting in MC@NLO-type
simulations led to the recommendation that all MEC be switched off, 
by means of:
\begin{verbatim}
SpaceShower:MEcorrections = off
TimeShower:MEcorrections = off
\end{verbatim}
An undesirable by-product of these settings is the absence of
MEC for decays. However, we have seen that even with the public
versions of \pye (starting with \pye.219), this can be avoided by setting:
\begin{verbatim}
SpaceShower:MEcorrections = off
TimeShower:MEcorrections = on
TimeShower:MEextended = off
\end{verbatim}
in the majority of the cases of interest including, but not limited
to, the processes that feature a $t\bt$ final state; both values of
\texttt{TimeShower::MEafterFirst} are acceptable. 

We stress again that this is not a clean separation between production
and decay MEC in the context of a generic process, and therefore the
above setting recommendations must not be seen as universal. As a 
cautionary tale, consider $Z$-mediated dijet production,
$q\bq\to Z^\star\to q\bq$. This process is seen by \pye as the production
of a colour singlet followed by its decay, rather than as a unique
$2\to 2$ production process. As such, exactly-matching matrix elements
exist for both production and decay ($Z\leftrightarrow q\bq g$), which
would not be the case had the $(2\to 2)$-process picture been adopted.
It follows that the value of \texttt{MEextended} is irrelevant in
this case, and MEC are applied or not depending solely on the value
of \texttt{MEcorrections}. Thus, if the latter is equal to \texttt{on},
one has (does not have) double counting if the process is generated
by \amcs as a $2\to 2$ ($2\to 1$ followed by tree-level $Z$ decay).
The bottom line is that, since avoiding any double counting is of 
paramount importance, in case of doubt the user is encouraged to 
contact the \pye authors.

\section{Monte Carlo event samples\label{sec:samples}}
We now proceed to evaluate the impact of MEC in resonance decays in
realistic NLO+PS-accurate MC simulations of relevance for LHC phenomenology.
For this, we consider different MC samples for the production of stable
top-quark pairs in proton-proton collisions at $\sqrt{s}=13$~TeV.

The first set of samples is generated using the \amcs~\cite{Alwall:2014hca}
program at the leading- and the next-to-leading-order (the latter with MC@NLO
matching) in the strong coupling constant. The renormalization and factorization
scales are set equal to the sum of the transverse energies of the final-state
particles at the hard-process level, namely the top, the antitop, and when
present a light quark or a gluon. The top quarks are subsequently decayed 
either leptonically or semileptonically (by including
only electrons and muons), and preserving the tree-level spin correlations, 
with MadSpin~\cite{Artoisenet:2012st}.  A second sample is generated using
standalone \pye.309 (thus, at the LO), with scales set equal to the
geometric mean of the squared transverse masses of the two outgoing particles.
The last sample is generated at the NLO with the HVQ~\cite{Frixione:2007nw}
program in \pwgb-V2~\cite{Frixione:2007vw,Alioli:2010xd}. In \pwgb the scales
are set equal to $\sqrt{m_t^2+\pt^2}$, with $m_t$ and $\pt$ the top quark 
mass and transverse momentum, respectively, evaluated by employing the 
underlying Born configuration; the $h_{\mathrm{damp}}$ parameter is set 
equal to $m_t$. The top decay is performed at the tree-level, and including
the effect of spin correlations, using the approach of
ref.~\cite{Frixione:2007zp} as is implemented in the \pwgb internal routines.
In all samples the NNPDF31\_nnlo\_as\_0118~\cite{NNPDF:2017mvq} parton
distributions are used, and the value of the top quark mass is set equal to
$m_t=172.5$~GeV. Both the \amcs LO sample and the \pye standalone one
are normalized, prior to any final-state cuts, to the NLO total cross 
section as is predicted by \amcs.

All of the predictions are then interfaced to \pye.309 (employed with the
Monash tune~\cite{Skands:2014pea}) to include the effects of parton showering,
multiple parton interactions and underlying event.  For the standalone \pye
and for the \pwgb samples, MEC are included wherever available, for both
production and decay.  For each of the \amcs LO and NLO event sets, we shower 
the events twice. Namely, we use the current recommended MEC settings,
in which matrix-element corrections are completely switched off, as well as
the newer settings presented in sect.~\ref{sec:py8MEC}, in which MEC are 
included only in the decays. Finally, the showered particle-level events 
are passed through Rivet~\cite{Bierlich:2019rhm}-based 
analyses that implement the observables of interest.

\section{Phenomenological results\label{sec:pheno}}
The inclusion of MEC to resonance decays is in general expected to affect
the kinematics of the reconstructed decay products.
We shall show in this section a few illustrative observables that document
the extent of these effects. In particular, the \pye LO+PS and \pwgb+\pye 
NLO+PS predictions that include MEC to production and decays are compared 
with \amcs NLO+PS predictions where MEC are either switched off (the current
default), or included only for the decay (the settings proposed here). 
Since the focus of this work is the impact of MEC, no effort is made
to simulate hard radiation {\em in production} in a manner which is as 
mutually consistent as possible across the various programs we employ.
We note that the corresponding phase-space region is in any case 
better described by merged approaches, which also feature smaller
systematics w.r.t.~those of matched simulations. 
In order to facilitate the disentangling of the effects due to 
decay MEC from those due to hard radiation in the context of simulations
stemming from exactly the same assumptions, in appendix~\ref{sec:LOapp}
we present comparisons between the LO- and NLO-accurate predictions of 
\amcs, for the same observables as those considered in this section.

All of the figures of this section have the same layout, namely consist of a
main frame and a lower inset. In the main frame the four predictions (and
possibly the experimental data) are displayed in absolute value, as blue
(\amcs with decay MEC), green (\amcs without MEC), orange (\pwgb+\pye), and
red (\pye) histograms. The lower inset shows the ratios of the latter three 
predictions (and possibly of the data) over the one for \amcs with decay MEC, 
using the same colour patterns as in the main frame.
For ease of reading the plots, the labels indicate,
where relevant, which kind of MEC are applied: MEC in decay 
(${\rm MEC}^{\rm dec}$) for \amcs, and MEC in both production and decay 
(${\rm MEC}_{\rm prod}^{\rm dec}$) for \pwgb and \pye standalone. 

\begin{figure}[!htb]
 \centering
  \includegraphics[width=.495\textwidth]{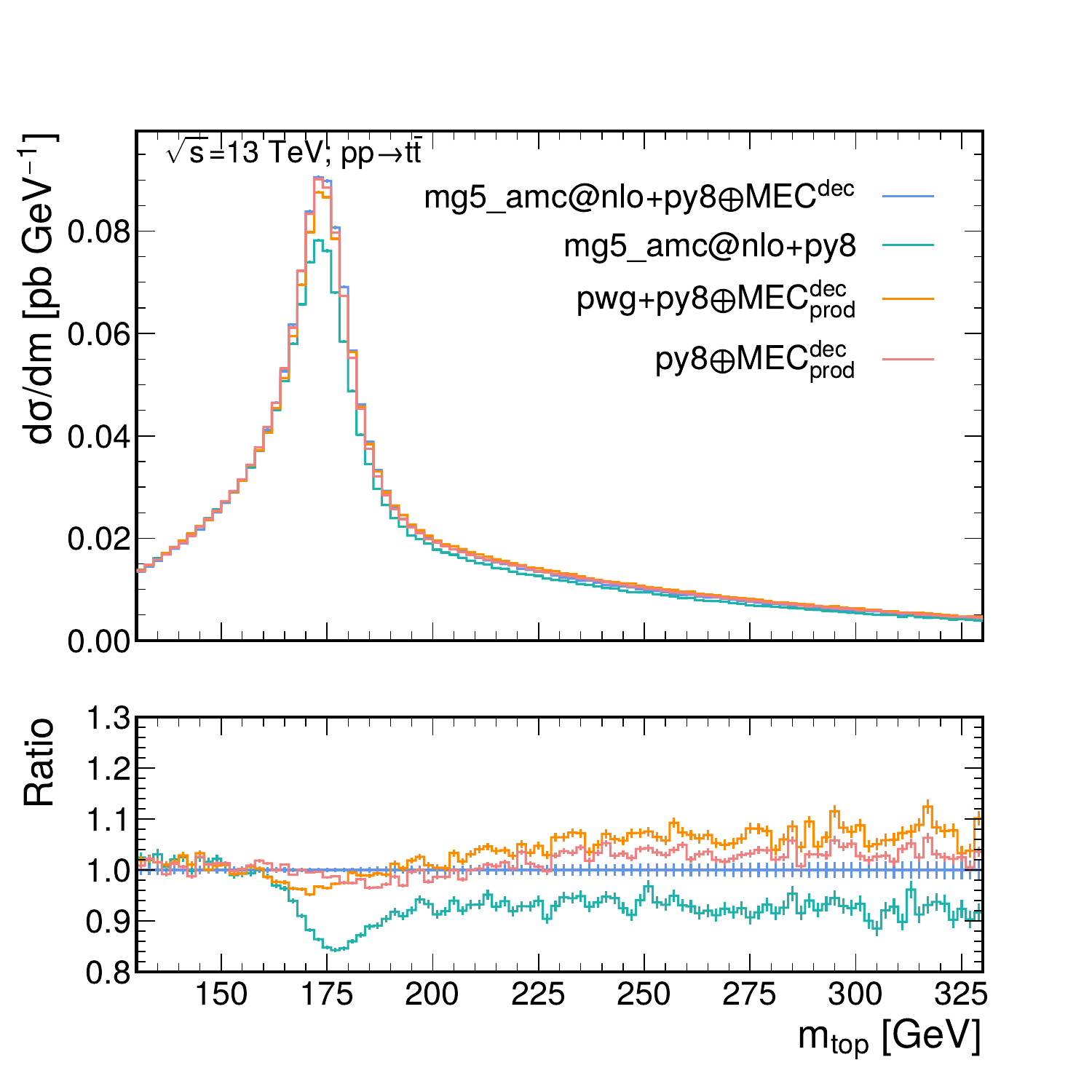}
  \includegraphics[width=.495\textwidth]{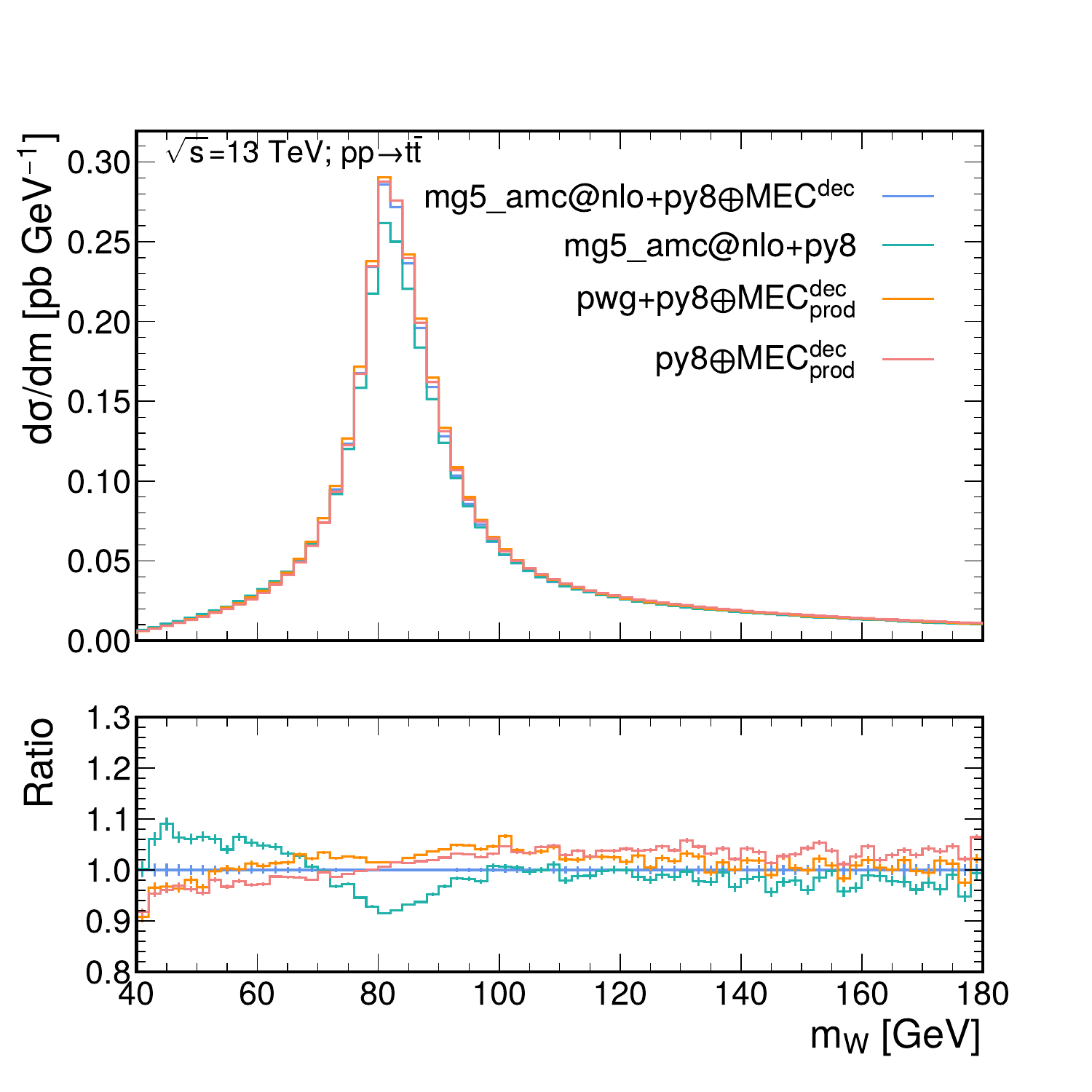}
  \caption{\label{fig:topWmass} The distribution of the reconstructed top
    quark (left panel) and $W$-boson (right panel) mass in semileptonic
    $t\bar{t}$ events.}
\end{figure}

For the case of the semileptonic channel, we can expect differences
in the reconstructed hadronic $W$-boson and top quark masses.
These two distributions are shown in fig.~\ref{fig:topWmass}.
The MEC in decay lead to a difference of about 10\% (20\%) in the 
reconstructed $W$-boson (top-quark) mass shape in the vicinity of the 
resonance peak. After inclusion of MEC, the three generators considered are
in a better than 5\% agreement with each other around the resonances peak,
where the validity of the narrow-width approximation is good, and the
impact of production hard radiation is negligible. As one can see
from fig.~\ref{fig:topWmass-lo}, hard radiation does have a visible
(but still moderate) impact at large invariant masses, more pronounced 
in the case of the reconstructed top mass than for the $W$ mass, which 
explains the small residual differences among our three benchmark results 
in that region.
%%%%%%%%%%%%%%%%%%%%%%%%%%%%%%%%%%%%%%%%%%%%%%%%%%%%%%%%%%%%%%%%%%%%%%%%%
%%%%%%%%%%%%%%%%%%%%%%%%%%%%%%%%%%%%%%%%%%%%%%%%%%%%%%%%%%%%%%%%%%%%%%%%%

%%%%%%%%%%%%%%%%%%%%%%%%%%%%%%%%%%%%%%%%%%%%%%%%%%%%%%%%%%%%%%%%%%%%%%%%%
\begin{figure}[!htb]
  \centering
  \includegraphics[width=.495\textwidth]{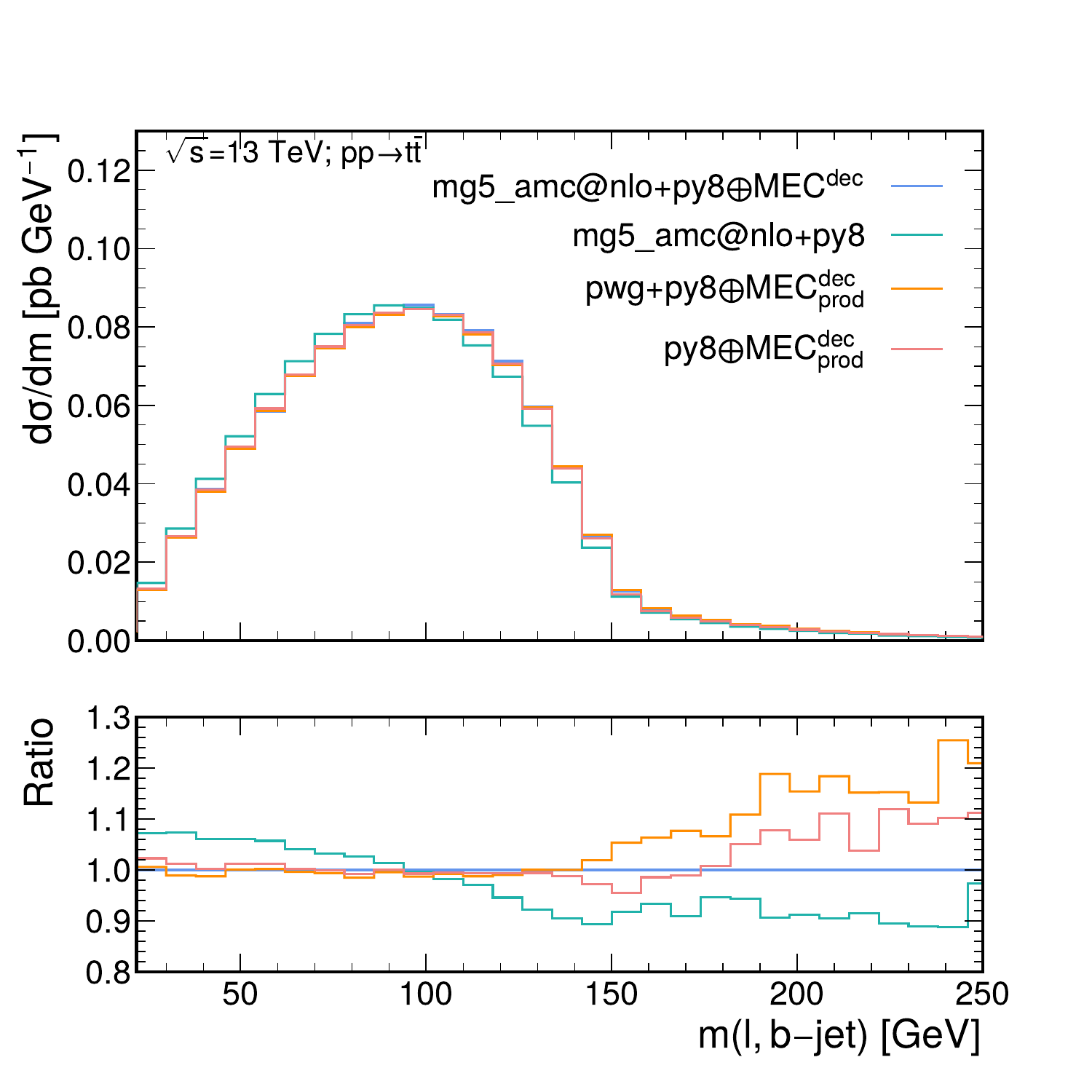}
  \includegraphics[width=.495\textwidth]{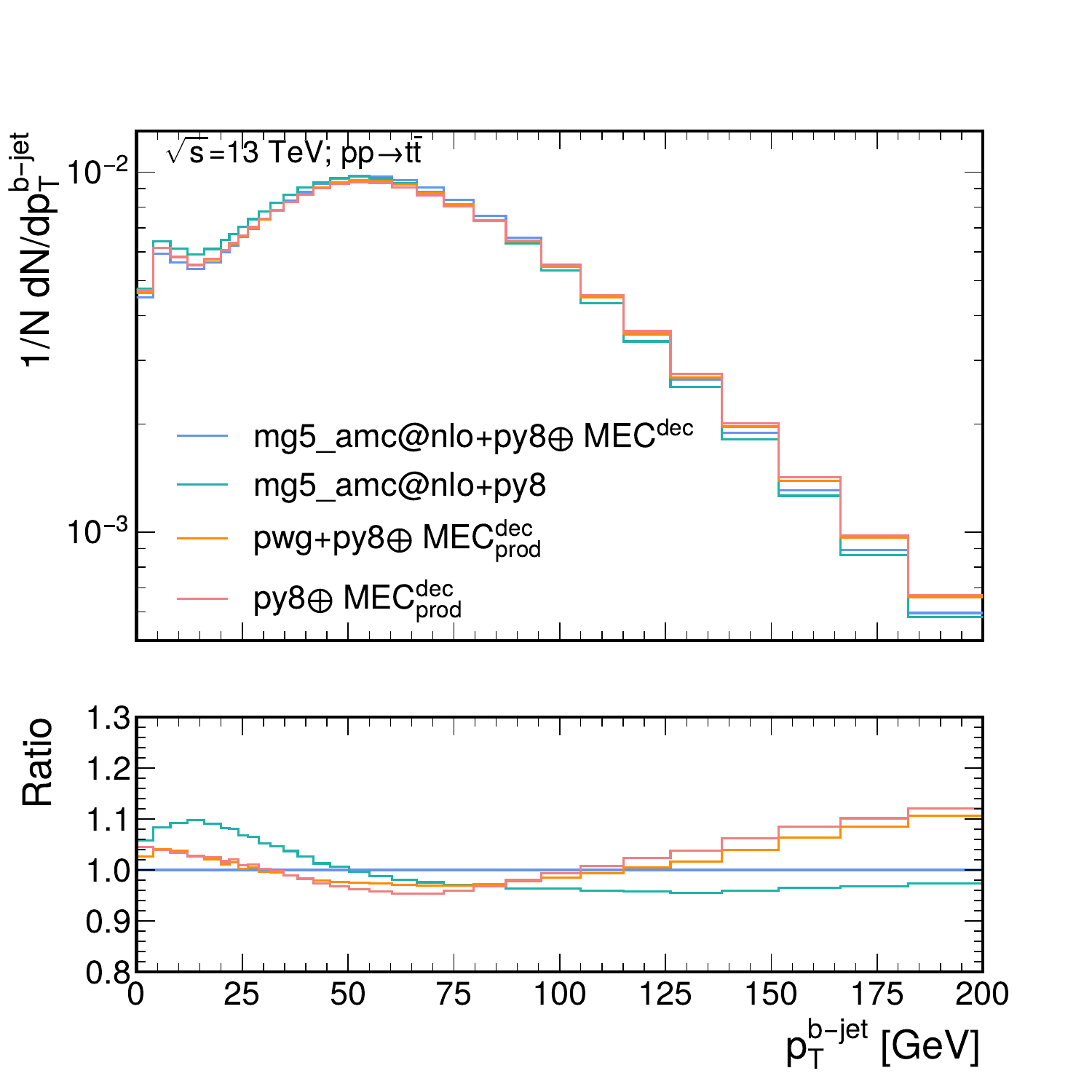}
  \caption{\label{fig:lepbjetpt} The distribution in the invariant mass of the
    (lepton,$b$-jet) pair (left panel) and the leading $b$-jet transverse
    momentum (right panel) in dileptonic $t\bar{t}$ events.
}
\end{figure}
%%%%%%%%%%%%%%%%%%%%%%%%%%%%%%%%%%%%%%%%%%%%%%%%%%%%%%%%%%%%%%%%%%%%%%%%%
The invariant mass of the (lepton,$b$-tagged jet) pair,
\mbox{$m(l,b_\mathrm{jet})$} and the leading $b$-tagged jet transverse
momentum are shown in fig.~\ref{fig:lepbjetpt} for dileptonic $t\bt$ events.  
The \mbox{$m(l,b_\mathrm{jet})$} distribution exhibits a kinematic edge around
$\sqrt{m_t^2-m_W^2}\sim 150$~GeV, sensitive to the value of the top mass. 
We observe that decay MEC shift the position of the peak, and after their
inclusion in \amcs this simulation and the \pwgb one (and to a good extent
that of \pye as well) agree fairly well with each other around and below this
peak. At larger \mbox{$m(l,b_\mathrm{jet})$} values the detailed description
of the production mechanism, which is treated differently in the three codes,
becomes important, and we observe relative differences of up to 30\% in this
region. Inspection of the left panel of fig.~\ref{fig:lepbjetpt-lo} confirms
that the vast majority of these discrepancies are indeed due to hard
radiation, with a residual ${\cal O}(5\%)$ effect stemming from decay MEC. The
impact of decay MEC is also evident at small values of the $b$-jet transverse
momentum (for $\pt$ smaller than about 50~GeV); thus, in this region decay MEC
improve the agreement between the three generators. However, at variance with
\mbox{$m(l,b_\mathrm{jet})$}, for the present observable the separation
between hard-radiation and decay-MEC effects is less clear-cut; this can be
best understood by looking at the right panel of
fig.~\ref{fig:lepbjetpt-lo}. Having said that, it is at large $\pt$ that the
impact of hard radiation is larger than that of decay MEC; hence, the
inclusion of decay MEC in \amcs does not help reduce significantly the 
${\cal O}(10\%)$ discrepancies between \amcs, \pwgb, and \pye.

\begin{figure}[!htb]
  \centering
  \includegraphics[width=.495\textwidth]{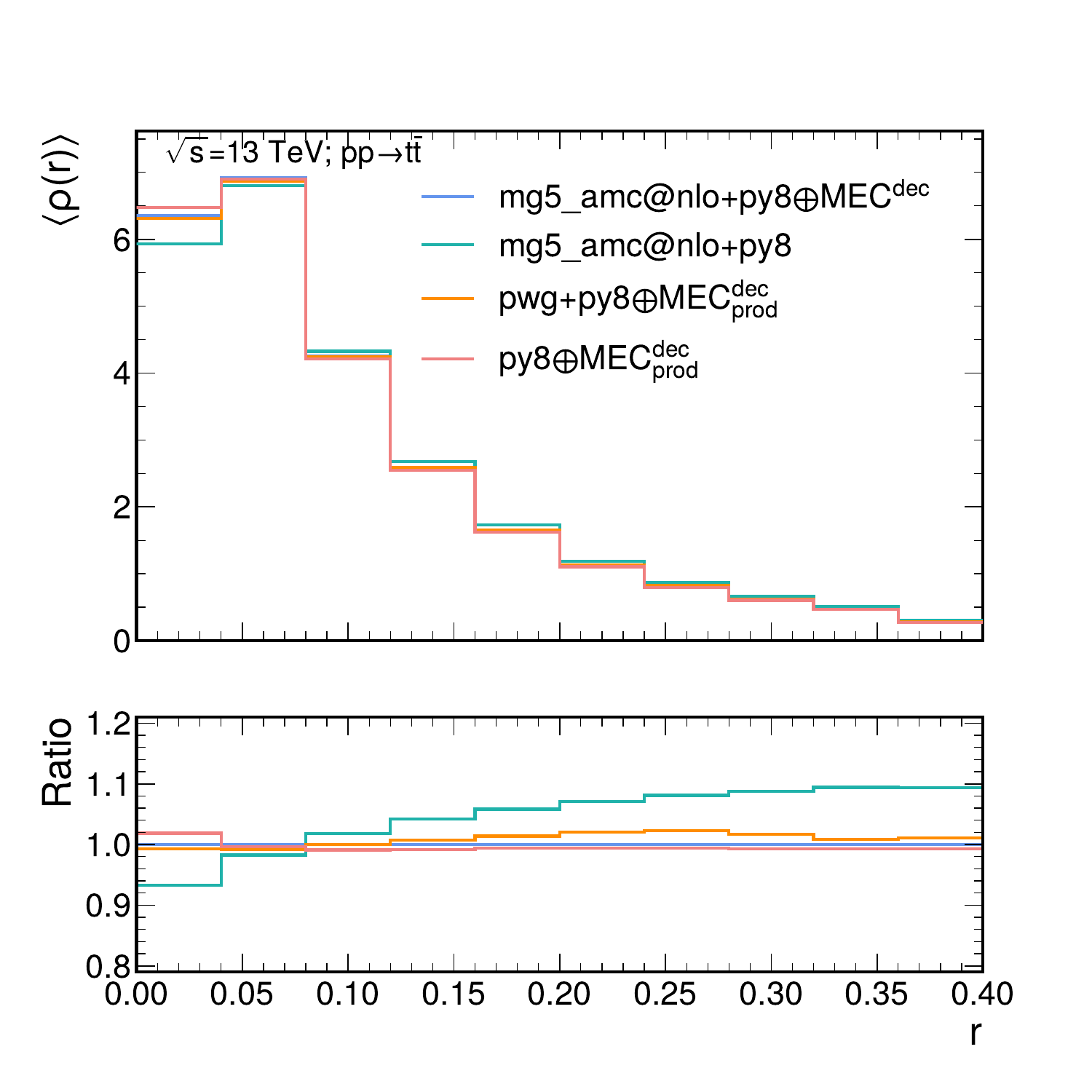}
  \includegraphics[width=.495\textwidth]{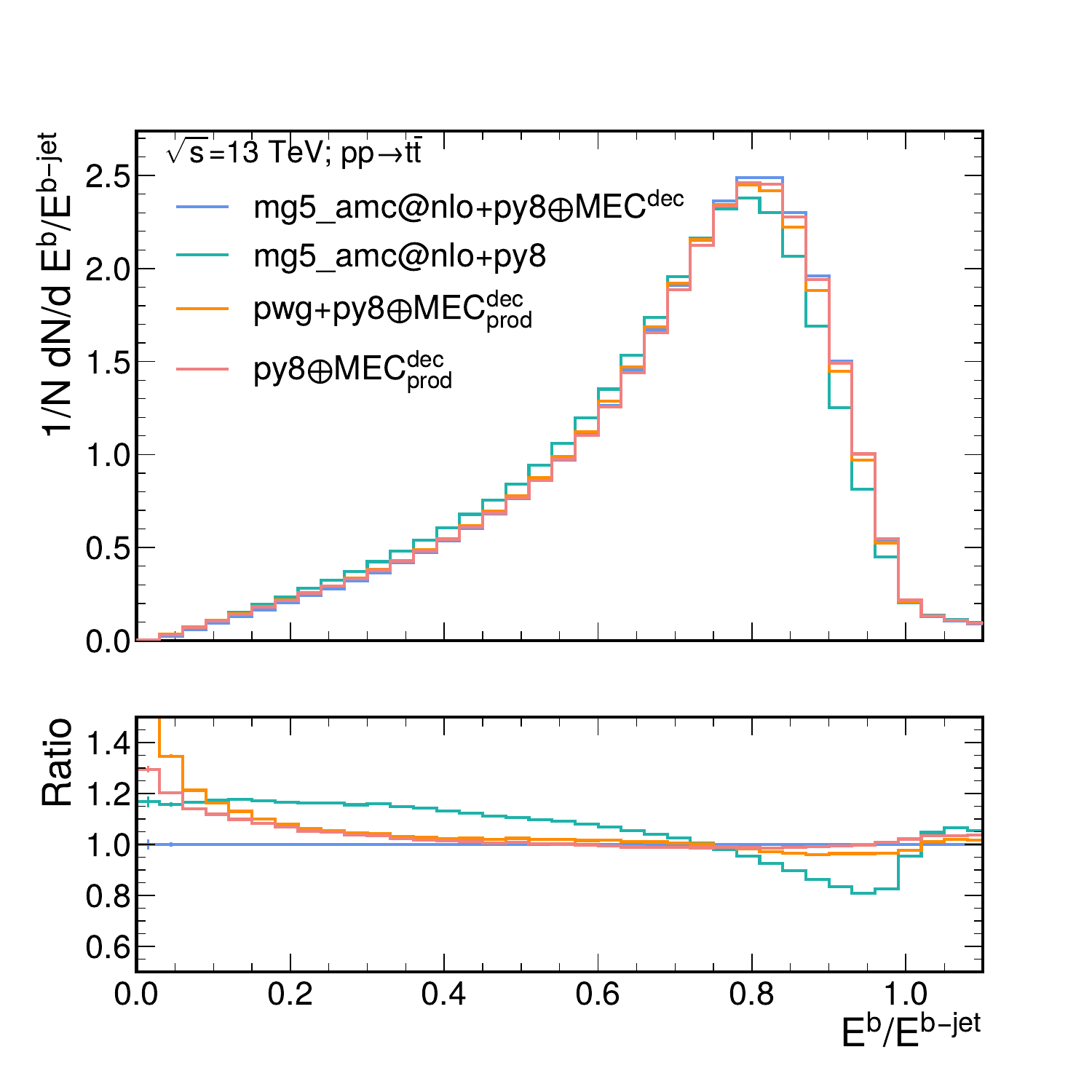}
  \caption{\label{fig:bjet} The differential $b$-jet shape distribution (left
    panel) and the scaled energy fraction of the $B$-hadron (right panel) in
    semileptonic $t\bar{t}$ events.  }
\end{figure}

Since decay MEC modify the kinematic properties of the $b$-quark and of the
$B$-hadron resulting from the hadronization of the former, they also have an
impact on the radiation pattern inside its corresponding $b$-tagged jet. In
order to illustrate this, in fig.~\ref{fig:bjet} we consider the distribution
of the $b$-jet profile, and of the scaled $B$-hadron energy spectrum.  The
$b$-jet profile $r$, a.k.a.~$b$-jet shape, is defined as the average fraction of
the jet transverse energy that lies inside an inner cone of radius $r<R$,
with $R$ the jet-radius parameter; jets are defined according to the
anti-$\kt$ algorithm~\cite{Cacciari:2008gp}, with parameters that
depend on the specific analysis one considers. The scaled $B$-hadron energy, 
defined as the ratio of the $B$-hadron energy over the energy of the $b$-jet 
containing it (with energies defined in the lab frame), is a proxy for the
longitudinal $b$-quark fragmentation function.  The decay MEC are found to
narrow the distribution of energy around the $b$-jet axis (i.e.~there are
comparatively more events at small $r$ values) , and to shift the
peak of the scaled $B$-hadron energy spectrum to higher values.  
The effect of decay MEC is significant in the whole $r$ range considered,
while that of hard radiation is negligible (see the left panel of
fig.~\ref{fig:bjet-lo}): thus, after their inclusion, the agreement among 
the \amcs, \pwgb, and \pye simulations is of ${\cal O}(2\%)$. The radiation
pattern is more complicated in the case of the scaled $B$-hadron energy
(see the right panel of fig.~\ref{fig:bjet-lo}), with residual non-negligible
hard-radiation effects at small values; still, in the bulk of the distribution
the agreement among our three benchmark results is at the same level as for
the jet energy profile.
%%%%%%%%%%%%%%%%%%%%%%%%%%%%%%%%%%%%%%%%%%%%%%%%%%%%%%%%%%%%%%%%%%%%%%%%%
%%%%%%%%%%%%%%%%%%%%%%%%%%%%%%%%%%%%%%%%%%%%%%%%%%%%%%%%%%%%%%%%%%%%%%%%%

%%%%%%%%%%%%%%%%%%%%%%%%%%%%%%%%%%%%%%%%%%%%%%%%%%%%%%%%%%%%%%%%%%%%%%%%%
\begin{figure}[!htb]
  \centering
  \includegraphics[width=.495\textwidth]{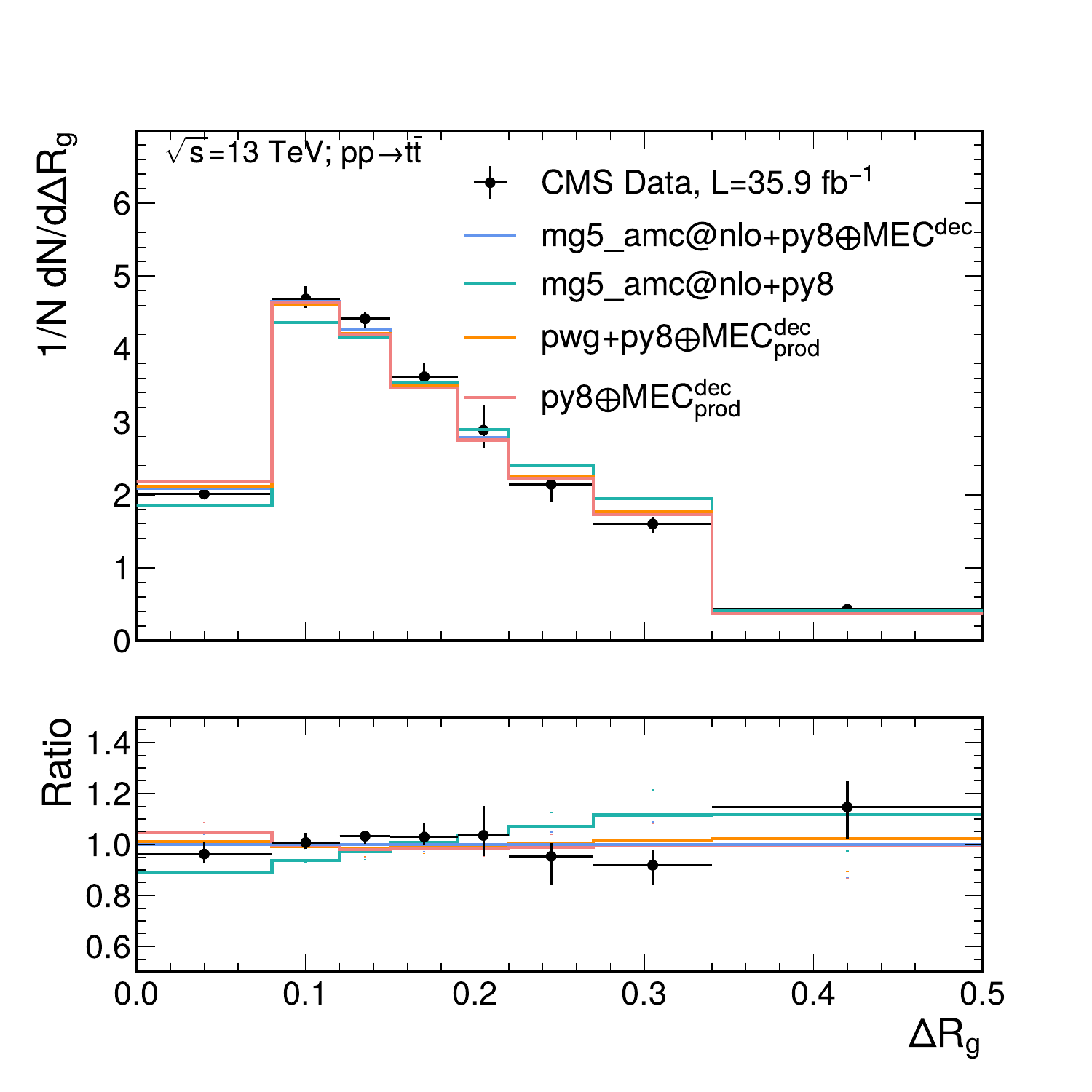} %CMS_2018_I1690148/d41-x01-y02
  \includegraphics[width=.495\textwidth]{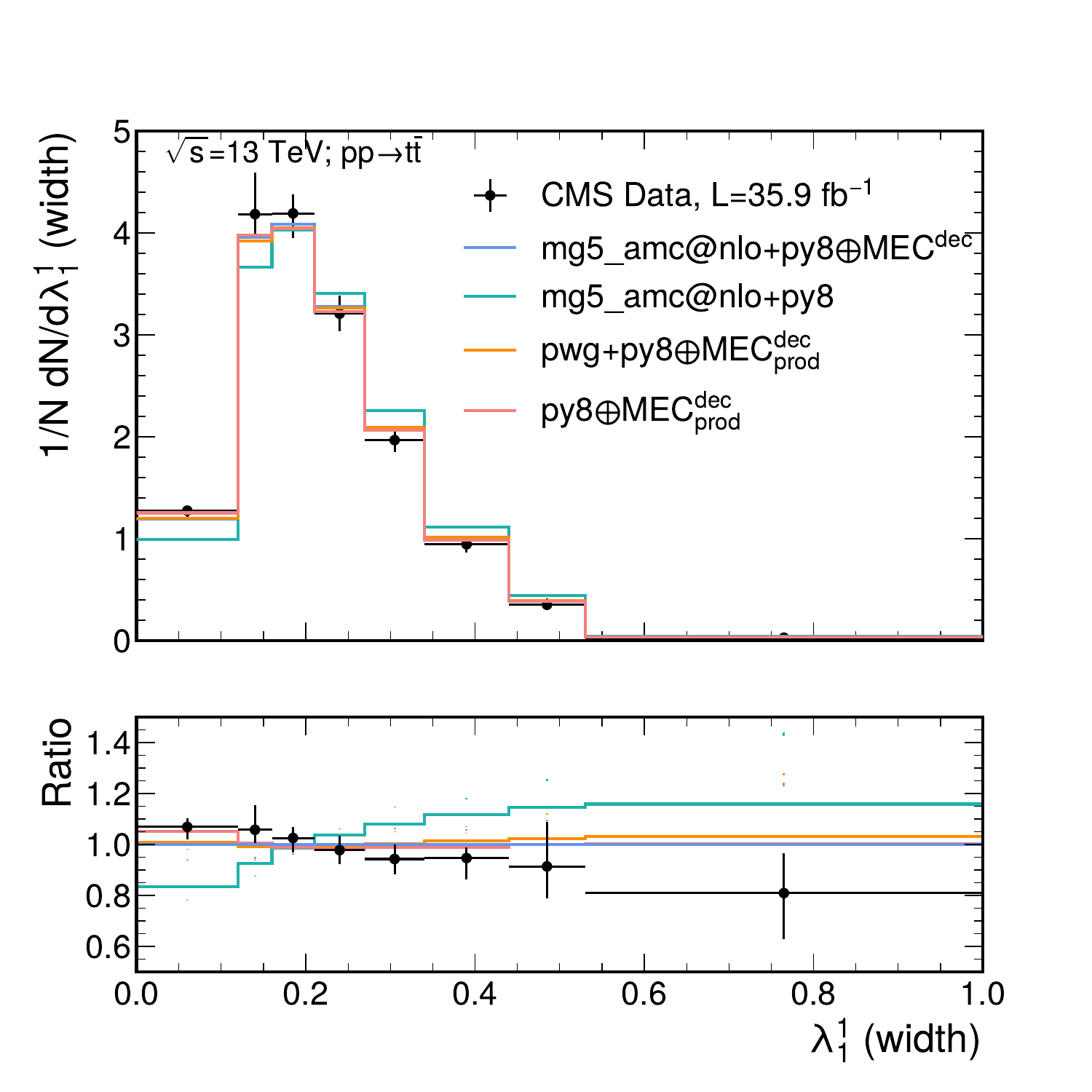} %CMS_2018_I1690148/d37-x01-y02
  \caption{\label{fig:cmsbjets} The groomed subjet distance (left panel) and
    the jet width (right panel) distribution as reconstructed from charged
    particle tracks in semileptonic $t\bar{t}$ events, compared to measured
    data by the CMS Collaboration~\cite{CMS:2018ypj}.  }
\end{figure} 
%%%%%%%%%%%%%%%%%%%%%%%%%%%%%%%%%%%%%%%%%%%%%%%%%%%%%%%%%%%%%%%%%%%%%%%%%
We finally compare our predictions to experimental data of jet substructure in
$t\bt$ semileptonic events that use charged-particle tracks from the CMS
collaboration~\cite{CMS:2018ypj}. In fig.~\ref{fig:cmsbjets} we consider two
observables that are sensitive to the radiation pattern inside $b$-tagged
jets: the groomed subjet distance, $\Delta R_g$, and the jet width,
$\lambda_1^1$. As is clearly demonstrated by fig.~\ref{fig:cmsbjets-lo},
these observables are essentially insensitive to hard radiation, and
decay MEC dominate their behaviour. Given what has been observed so
far, it is therefore not particularly surprising that the inclusion
of decay MEC in \amcs vastly improve its agreement with both \pwgb
and \pye results, whereas the previous MEC settings in \amcs led to
discrepancies of ${\cal O}(\pm 15\%)$ with these two codes. The
description of the data is also significantly improved.

\section{Conclusions\label{sec:concl}}
We have illustrated how matrix element corrections to resonance decays as
implemented in the \pye parton shower can be consistently included in
MC@NLO-type NLO+PS simulations in order to improve their phenomenological
accuracy, without resulting in any double counting. We have discussed the
impact of these corrections in a process of particular relevance for LHC
physics, namely the production of top quark pairs. We have also verified that
the same conclusions apply to the associated production of top-quark pairs and
a Higgs boson -- this must be expected, since this kind of effects are thought
to largely factorize w.r.t.~the hard process; we thus regard our findings as
to be universally valid.  We have found that decay MEC can have a relative
impact on the shape of distributions of up to 20\%. By comparing NLO+PS
predictions stemming from MC@NLO- and POWHEG-type matching with standalone
\pye ones we find that a significant reduction in the spread among the results
occurs if the MEC are included whenever possible in the various
simulations. We thus encourage the usage of the MEC settings proposed in this
paper for any practical applications, in order to improve the phenomenological
accuracy of MC@NLO-type simulations, and to reduce systematic uncertainties.

\vskip 2mm
\section*{Acknowledgments}
The authors thank Stefan Prestel for initiating a modification to the \pye 
code to allow control of matrix element corrections at different stages of 
the event evolution, as well as Paolo Nason, Simon Pl\"azter, and
Silvia Ferrario Ravasio for useful information. SF thanks the CERN TH 
division for the kind hospitality during the course of this work.

\paragraph{Funding information}
SM is supported by the Fermi Research Alliance, LLC under Contract
No. DE-AC02-07CH11359 with the U.S. Department of Energy, Office of Science,
Office of High Energy Physics.
SA is supported by the Helmholtz Association under the contract W2/W3-123.

\appendix
\section{\pye settings\label{sec:pyset}}
We report in this appendix the \pye settings used to obtain the various
predictions presented in this paper, and in particular the settings used 
to enable MEC to top decays in MC@NLO-type simulations.
Our numerical results are based on \pye.306.
No updates relevant to this study have occurred up to the current
release \pye.310. \\

\noindent 
All settings not listed take their default values, corresponding to the 
Monash tune based on the \texttt{NNPDF2.3 QCD+QED LO} parton distributions 
functions.

\subsection{\pye standalone}

\begin{verbatim}
Top:gg2ttbar = on
Top:qqbar2ttbar = on
\end{verbatim}

\subsection{\amcs}
\begin{verbatim}
SpaceShower:pTmaxMatch = 1
SpaceShower:pTmaxFudge = 1
TimeShower:pTmaxMatch = 1
TimeShower:pTmaxFudge = 1
TimeShower:globalRecoil = on
TimeShower:limitPTmaxGlobal = on
TimeShower:nMaxGlobalRecoil = 1
TimeShower:globalRecoilMode = 2
TimeShower:nMaxGlobalBranch = 1
TimeShower:weightGluonToQuark = 1
SpaceShower:MEcorrections = off
TimeShower:MEcorrections = on
TimeShower:MEextended = off
\end{verbatim}

\subsection{\pwgb}

\begin{verbatim}
POWHEG:nFinal = 2
POWHEG:veto = 1
POWHEG:pTdef = 1
POWHEG:emitted = 0
POWHEG:pTemt = 0
POWHEG:pThard = 0
POWHEG:vetoCount = 100
SpaceShower:pTmaxMatch = 2
TimeShower:pTmaxMatch = 2
\end{verbatim}

\section{Comparison with LO \amcs results\label{sec:LOapp}}
In this appendix we consider the same observables as in sect.~\ref{sec:pheno},
and compare the results obtained with \amcs+\pye by turning off and on
decay MEC. We do so both at the LO and the NLO accuracy; this allows one
to disentangle the effects of the MEC from those of hard radiation in
production in a more transparent manner w.r.t.~the comparison between
\amcs and either \pwgb or \pye, in view of the fact that all of the
simulations that appear here have identical settings for the short-distance 
cross sections.

All of the figures have the same layout, with a main frame and a lower
inset. In the main frame we show both NLO results (blue: with decay
MEC; green: without decay MEC) and LO results (orange: with decay
MEC; red: without decay MEC). Thus, the blue and green histograms 
here are the same as those in sect.~\ref{sec:pheno}. The insets present 
the ratios of the various predictions over the NLO-with-MEC one.

\begin{figure}[!htb]
\centering
  \includegraphics[width=.495\textwidth]{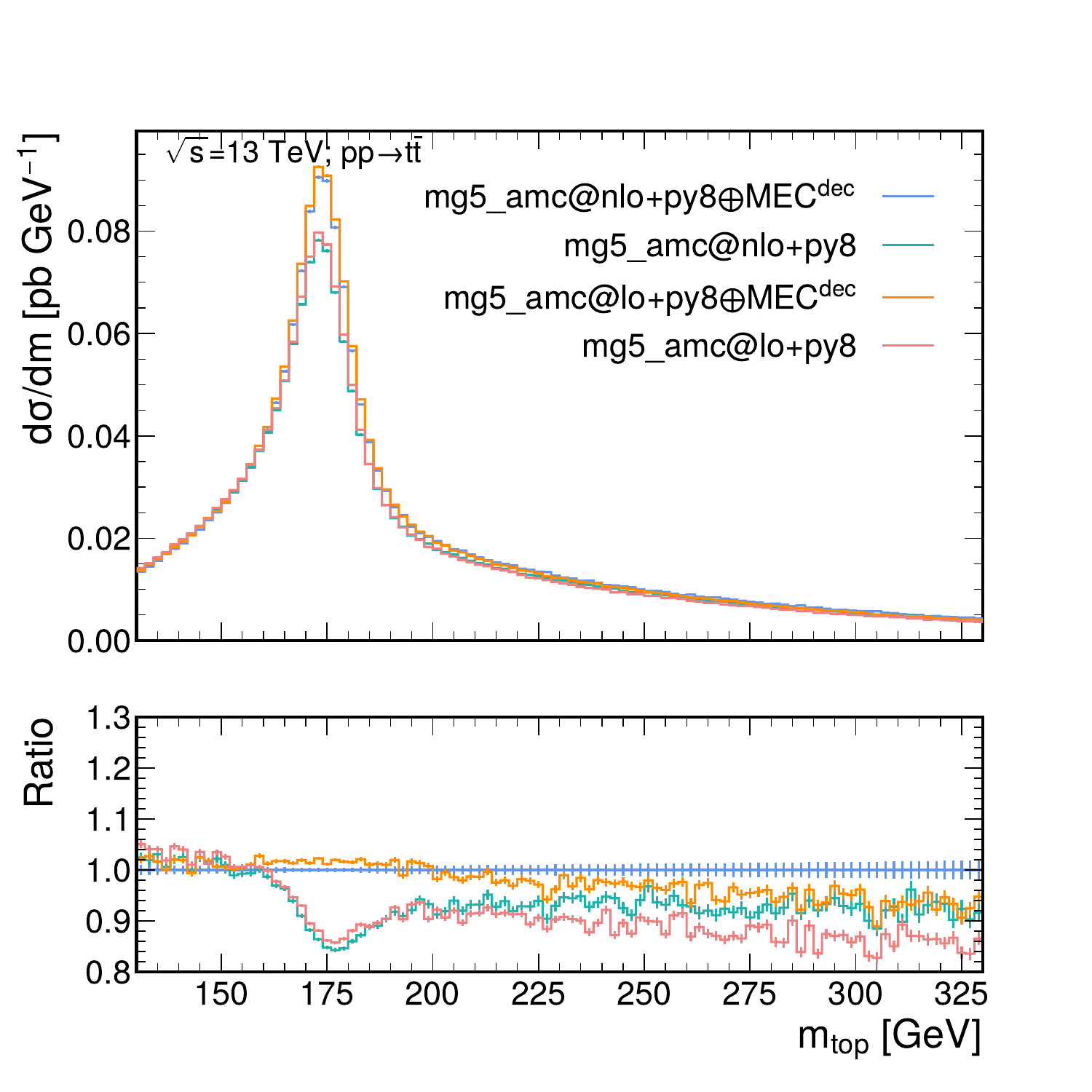}
  \includegraphics[width=.495\textwidth]{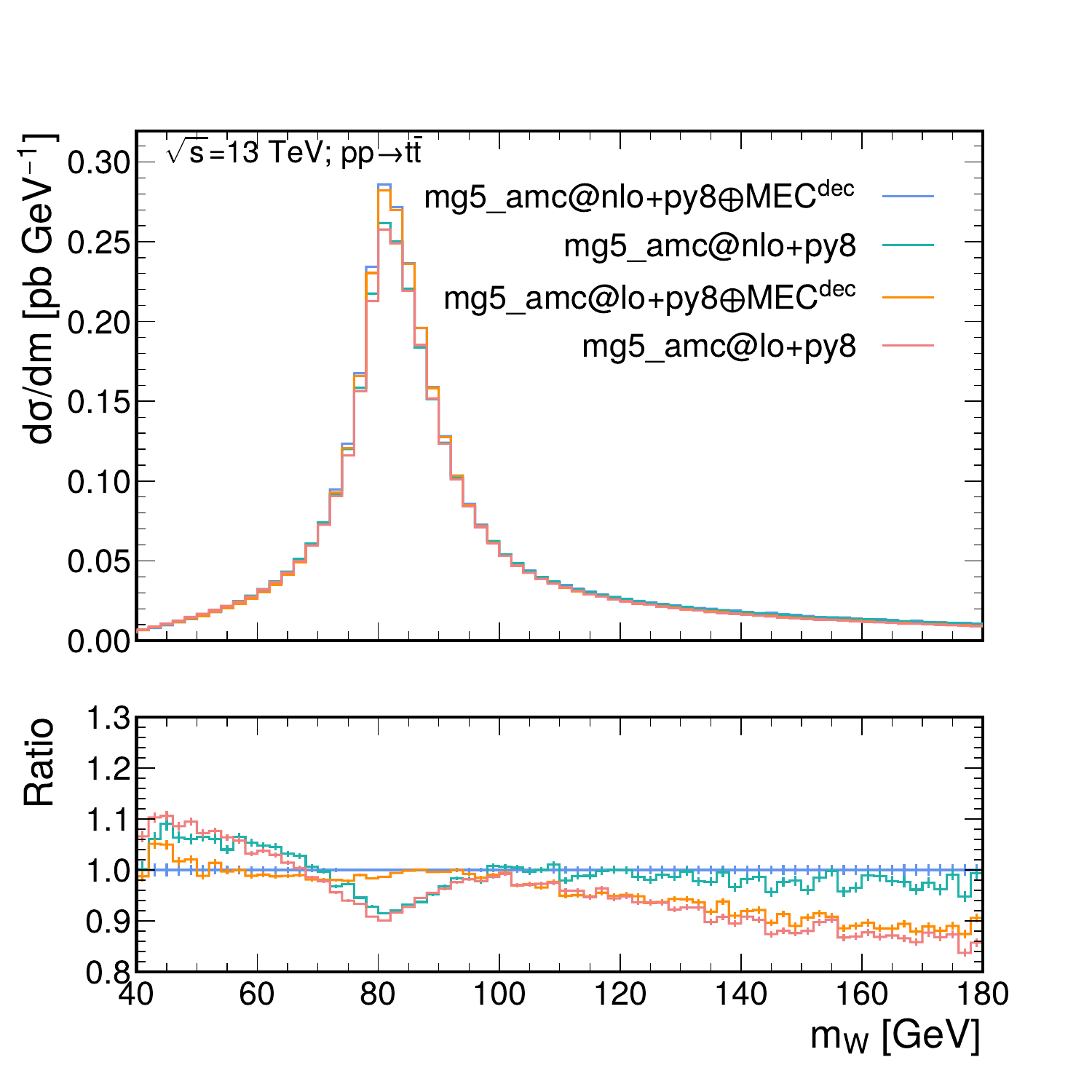}
  \caption{\label{fig:topWmass-lo} The distribution of the reconstructed top
    quark (left panel) and $W$-boson (right panel) mass in semileptonic
    $t\bar{t}$ events.  }
\end{figure}

\begin{figure}[!htb]
\centering
  \includegraphics[width=.495\textwidth]{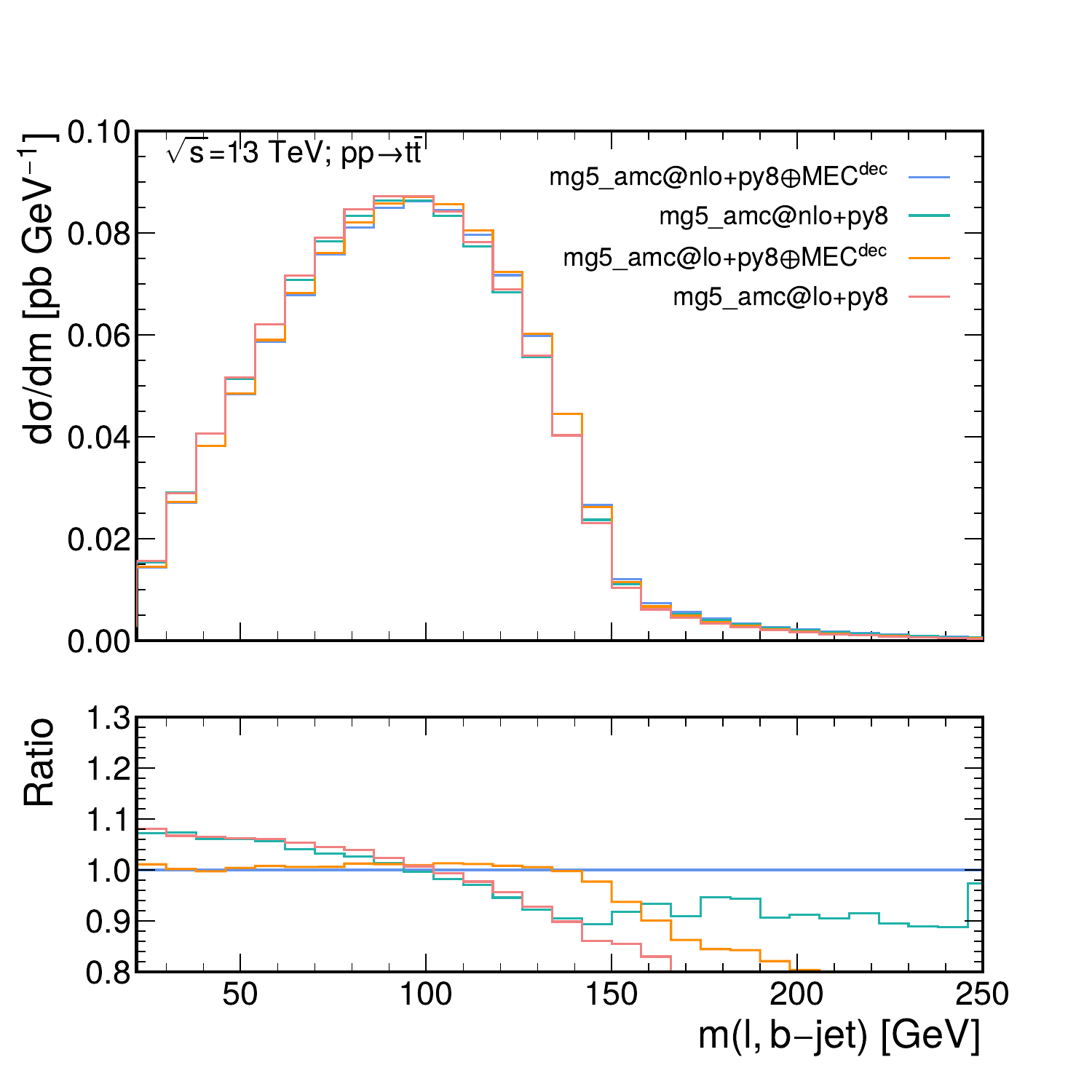}
  \includegraphics[width=.495\textwidth]{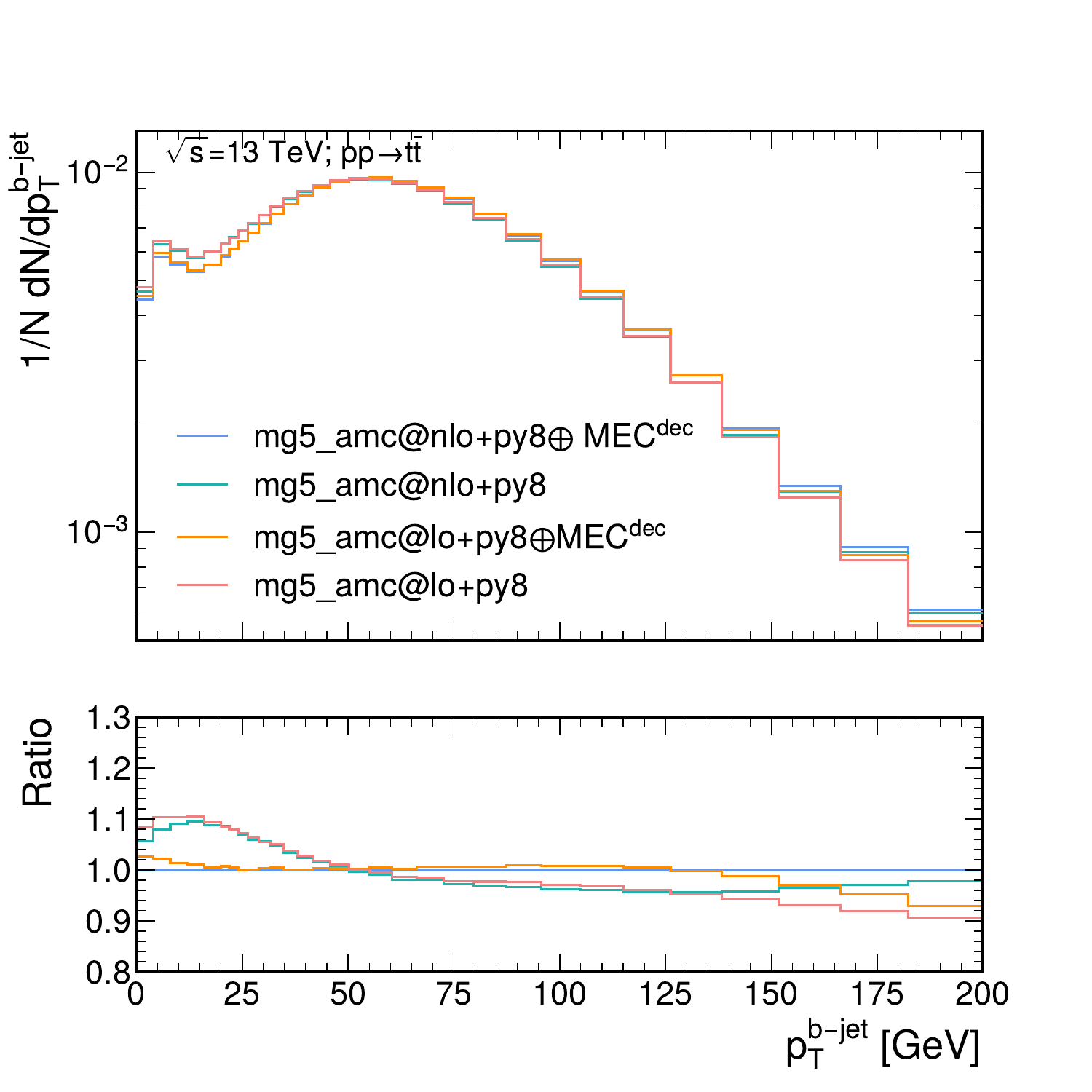}
 \caption{\label{fig:lepbjetpt-lo} The distribution in the invariant mass of
    the (lepton,$b$-jet) pair (left panel) and the leading $b$-jet transverse
    momentum (right panel) in dileptonic $t\bar{t}$ events.  }
\end{figure}

\begin{figure}[!htb]
\centering
  \includegraphics[width=.495\textwidth]{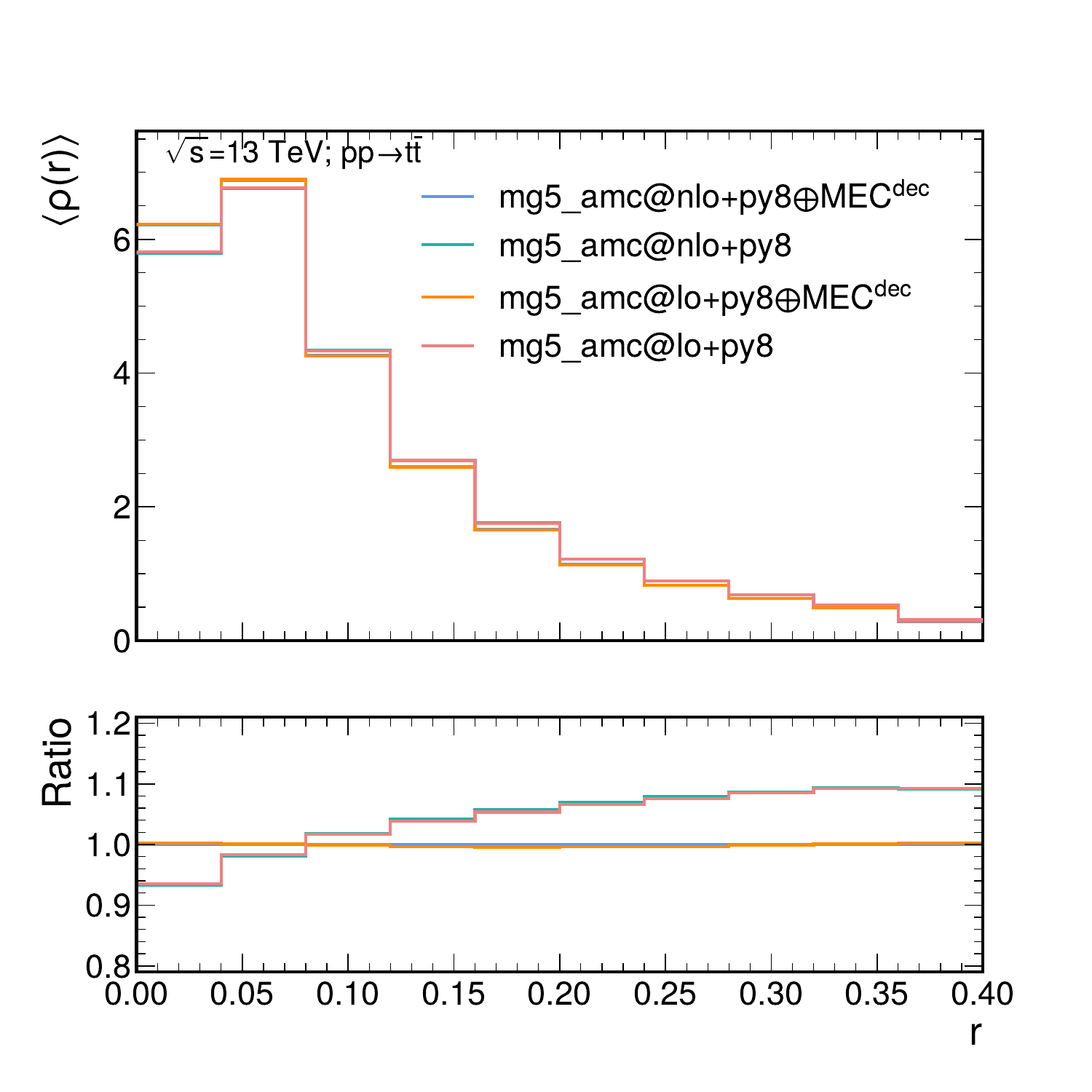}
 \includegraphics[width=.495\textwidth]{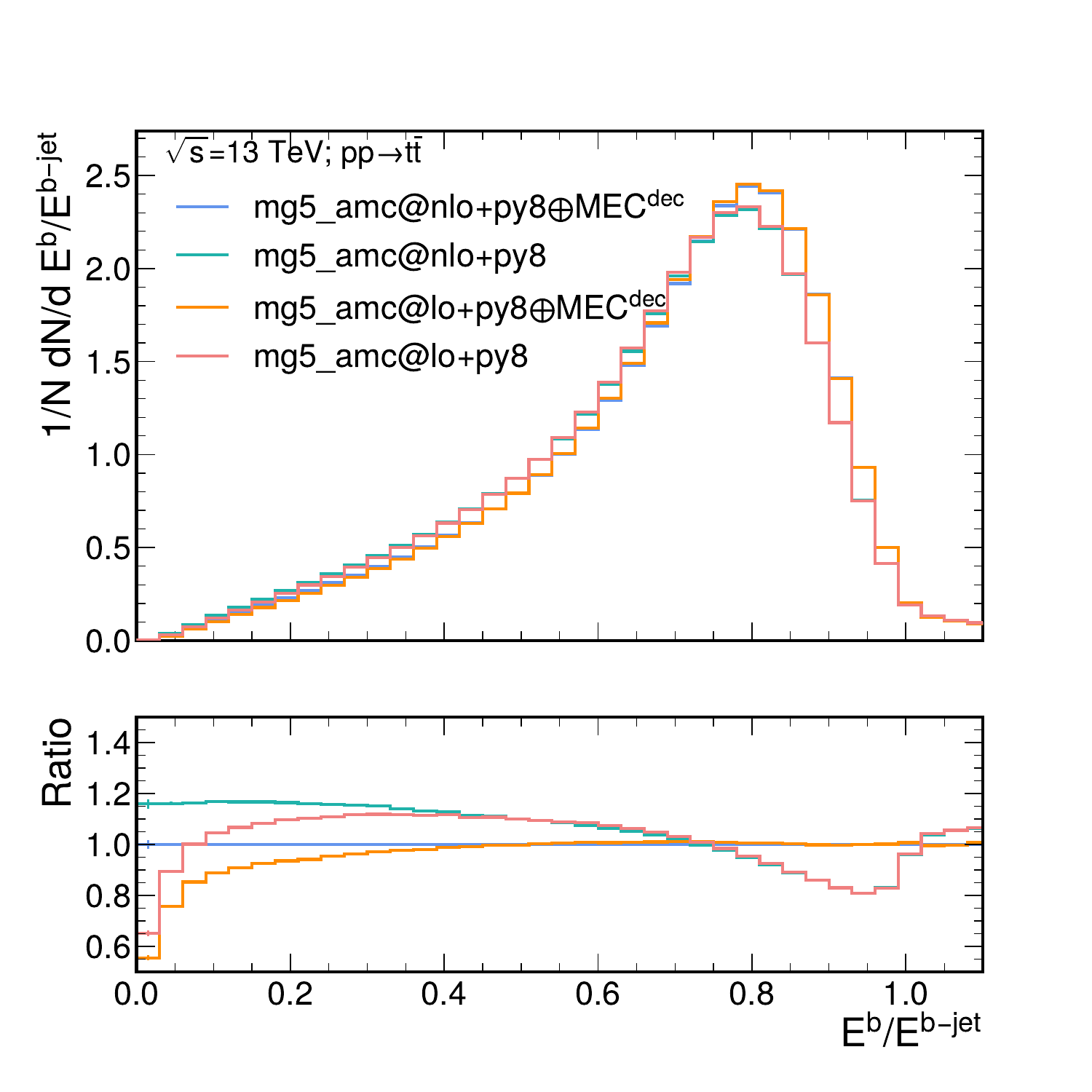}
  \caption{\label{fig:bjet-lo} The differential $b$-jet shape distribution
    (left panel) and the scaled energy fraction of the $B$-hadron (right
    panel) in semileptonic $t\bar{t}$ events.  }
\end{figure}

\begin{figure}[!htb]
\centering
  \includegraphics[width=.495\textwidth]{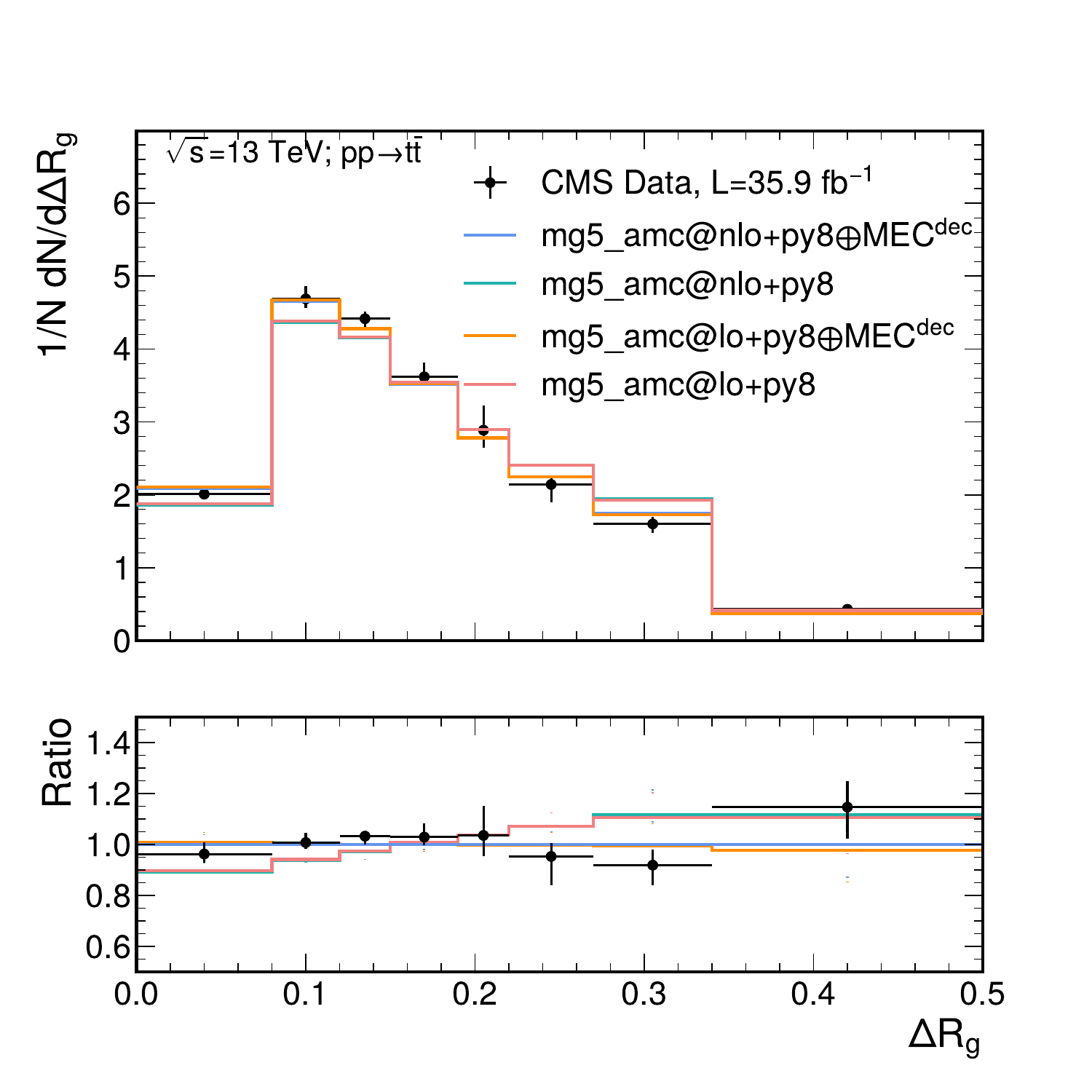}
  \includegraphics[width=.495\textwidth]{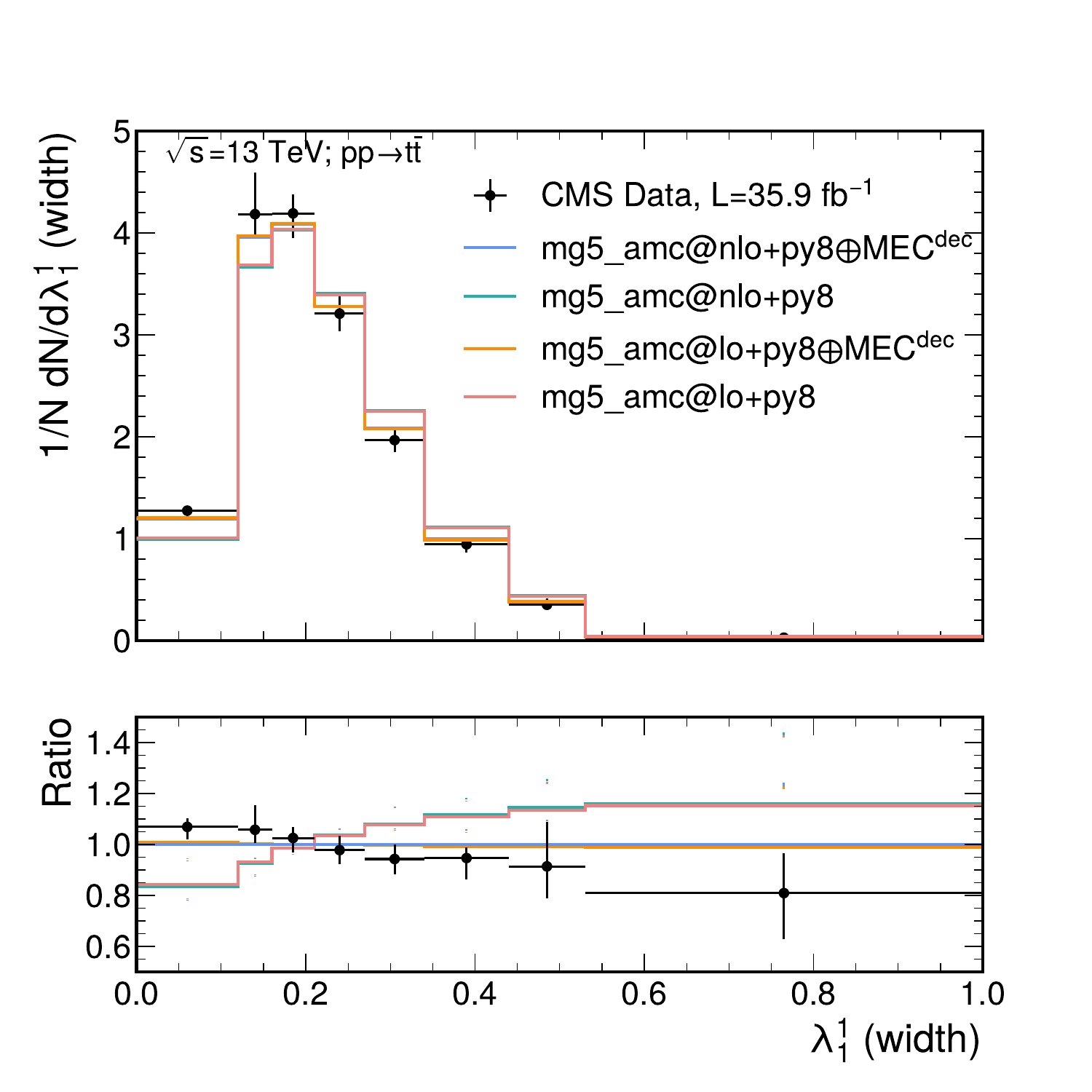}
  \caption{\label{fig:cmsbjets-lo} The groomed subjet distance (left panel)
    and the jet width (right panel) distribution as reconstructed from charged
    particle tracks in semileptonic $t\bar{t}$ events, compared to measured
    data by the CMS Collaboration~\cite{CMS:2018ypj}.  }
\end{figure}

Each of the figures in this appendix is a companion of a figure
in sect.~\ref{sec:pheno}; in particular, the observable displayed
in figs.~\ref{fig:topWmass-lo}, \ref{fig:lepbjetpt-lo}, \ref{fig:bjet-lo},
and~\ref{fig:cmsbjets-lo} are the same as those in \ref{fig:topWmass},
\ref{fig:lepbjetpt}, \ref{fig:bjet}, and~\ref{fig:cmsbjets}, respectively.
The reader is encouraged to compare systematically the figures of
this appendix with the corresponding ones in the main text.

In essence, in the plots shown here the differences between the blue and the
green histograms, and those between the orange and red histograms, indicate
that decay MEC have a non-negligible impact. Conversely, the differences
between the blue and the orange histograms, and those between the green and
red histograms, indicate that hard radiation has a non-negligible impact.

%%%%%%%%%%%%%%%%%%%%%%%%%%%%%%%%%%%%%%%%%%%%%%%%%%%%%%%%%%%%%%%%%%%%%%%%
%%%%%%%%%%%%%%%%%%%%%%%%%%%%%%%%%%%%%%%%%%%%%%%%%%%%%%%%%%%%%%%%%%%%%%%%
%%%%%%%%%%%%%%%%%%%%%%%%%%%%%%%%%%%%%%%%%%%%%%%%%%%%%%%%%%%%%%%%%%%%%%%%
%%%%%%%%%%%%%%%%%%%%%%%%%%%%%%%%%%%%%%%%%%%%%%%%%%%%%%%%%%%%%%%%%%%%%%%%
%%%%%%%%%%%%%%%%%%%%%%%%%%%%%%%%%%%%%%%%%%%%%%%%%%%%%%%%%%%%%%%%%%%%%%%%
%%%%%%%%%%%%%%%%%%%%%%%%%%%%%%%%%%%%%%%%%%%%%%%%%%%%%%%%%%%%%%%%%%%%%%%%
%%%%%%%%%%%%%%%%%%%%%%%%%%%%%%%%%%%%%%%%%%%%%%%%%%%%%%%%%%%%%%%%%%%%%%%%
%%%%%%%%%%%%%%%%%%%%%%%%%%%%%%%%%%%%%%%%%%%%%%%%%%%%%%%%%%%%%%%%%%%%%%%%

\section{Control of Matrix Element Corrections\label{app:c}}
The \pye setting \texttt{TimeShower:MEextended=off} disables the
equivalent-matrix-element FSR MEC for all parton emissions in the production
process.  As discussed in the text, consistency with MC@NLO-type matching
requires only that the MEC be disabled for the first emission.  However, in
the \pye shower, the MEC also provide the soft emission limit necessary for
particles with mass larger than or of the order of the soft emission scale.
These corrections lead to the dead-cone effect.

For the case of a heavy particle system radiating a gluon, such as $t\bar t
\to t\bar t g$, it can be argued that the MEC to heavy particle radiators
beyond the first emission, e.g. $t\bar t g \to t\bar t g g$, are unimportant
-- dipoles with the $g$ as the radiator will dominate over the mass suppressed
$t$ dipoles.  We have tested this explicitly by modifying the \pye shower to
disable the MEC for only a user-determined number of (QCD) emissions.  We
observe little or no impact of this choice on any of our numerical results.
To simplify the discussion and allow users to experiment with the impact of
the \texttt{MEextended} setting in a public version of \pye, we have not used
this new capability in our results.

This option will be publicly available in an upcoming \pye release.

\bibliography{MEC.bib}

\end{document}